\documentclass{article}

\usepackage{arXiv}

\usepackage[utf8]{inputenc} 
\usepackage[T1]{fontenc}    
\usepackage{hyperref}       
\usepackage{url}            
\usepackage{booktabs}       
\usepackage{amsfonts}       
\usepackage{nicefrac}       
\usepackage{microtype}      
\usepackage{lipsum}		
\usepackage{amsmath, amssymb}
\usepackage{multicol}
\usepackage{multirow}
\usepackage{caption}
\usepackage{xcolor}
\usepackage{blindtext}
\usepackage{graphicx}
\usepackage{doi}
\usepackage{enumitem}
\usepackage[affil-it]{authblk}
\usepackage{cleveref}
\crefname{figure}{Fig}{Figs}
\crefname{table}{Table}{Tables}
\usepackage{booktabs}
\usepackage{setspace} 
\usepackage{relsize}

\usepackage{xcolor}
\hypersetup{
    colorlinks,
    linkcolor={red},
    citecolor={blue},
    urlcolor={blue!80!black}
}

\title{DSA-NRP: No-Reflow Prediction from Angiographic Perfusion Dynamics in Stroke EVT}

\author[1,$\dagger$]{Shreeram Athreya}
\author[2,$\dagger$]{Carlos Olivares}
\author[3]{Ameera Ismail}
\author[4]{Kambiz Nael}
\author[2,3,5]{William Speier}
\author[1,2,3,5,6]{Corey W. Arnold}

\affil[1]{Department of Electrical and Computer Engineering, UCLA}
\affil[2]{Medical Informatics, UCLA}
\affil[3]{Department of Radiological Sciences, UCLA}
\affil[4]{Department of Radiology, UC San Francisco}
\affil[5]{Department of Bioengineering, UCLA}
\affil[6]{Department of Pathology and Laboratory Medicine, UCLA} \affil[$\dagger$]{Equal contribution \authorcr Corresponding author: Shreeram Athreya (\texttt{shreeram@ucla.edu})}

\date{}

\renewcommand{\headeright}{}
\renewcommand{\undertitle}{}

\begin{document}

\maketitle
 
\begin{abstract}
Following successful large-vessel recanalization via endovascular thrombectomy (EVT) for acute ischemic stroke (AIS), some patients experience a complication known as \emph{no-reflow}, defined by persistent microvascular hypoperfusion that undermines tissue recovery and worsens clinical outcomes. Although prompt identification is crucial, standard clinical practice relies on perfusion magnetic resonance imaging (MRI) within 24 hours post-procedure, delaying intervention. In this work, we introduce the first-ever machine learning (ML) framework to predict no-reflow immediately after EVT by leveraging previously unexplored intra-procedural digital subtraction angiography (DSA) sequences and clinical variables. Our retrospective analysis included AIS patients treated at UCLA Medical Center (2011–2024) who achieved favorable mTICI scores (2c or 3) and underwent pre- and post-procedure MRI. No-reflow was defined as a $>15\%$ reduction in relative cerebral blood volume or flow within the infarct core compared to the contralateral hemisphere. From DSA sequences (anteroposterior and lateral views), we extracted statistical and temporal perfusion features from the target downstream territory to train ML classifiers for predicting no-reflow. Our preliminary results demonstrate that this novel method outperformed a clinical-features baseline (AUROC: 0.9330 vs. 0.7768 ($p = 0.006$)), suggesting that real-time DSA perfusion dynamics may encode clinically relevant information related to microvascular integrity. This approach establishes a preliminary foundation for immediate, accurate no-reflow prediction, enabling clinicians to proactively manage high-risk patients without reliance on delayed imaging, though it warrants validation in larger, independent cohorts.
\end{abstract}


{S}{troke} is the third most prevalent cause of death worldwide, with acute ischemic stroke (AIS) being the most common type~\cite{saini_global_2021, kim_global_2024}. AIS occurs when blood flow to the brain is blocked by an occlusion, usually blood clots, in a cerebral artery, leading to a rapid deterioration of brain tissue from ischemia (tissue degeneration due to lack of oxygen) to infarction (tissue death resulting from prolonged ischemia). It is critical to restore blood flow as quickly as possible to prevent permanent brain damage~\cite{karonen_combined_1999}. Thrombolysis and endovascular thrombectomy (EVT) are the primary treatments for AIS, with EVT becoming more widely adopted due to its higher efficacy~\cite{badhiwala_endovascular_2015,ganesh_thrombectomy_2018}.

\begin{figure*}
    \centering
    \includegraphics[width=\textwidth]{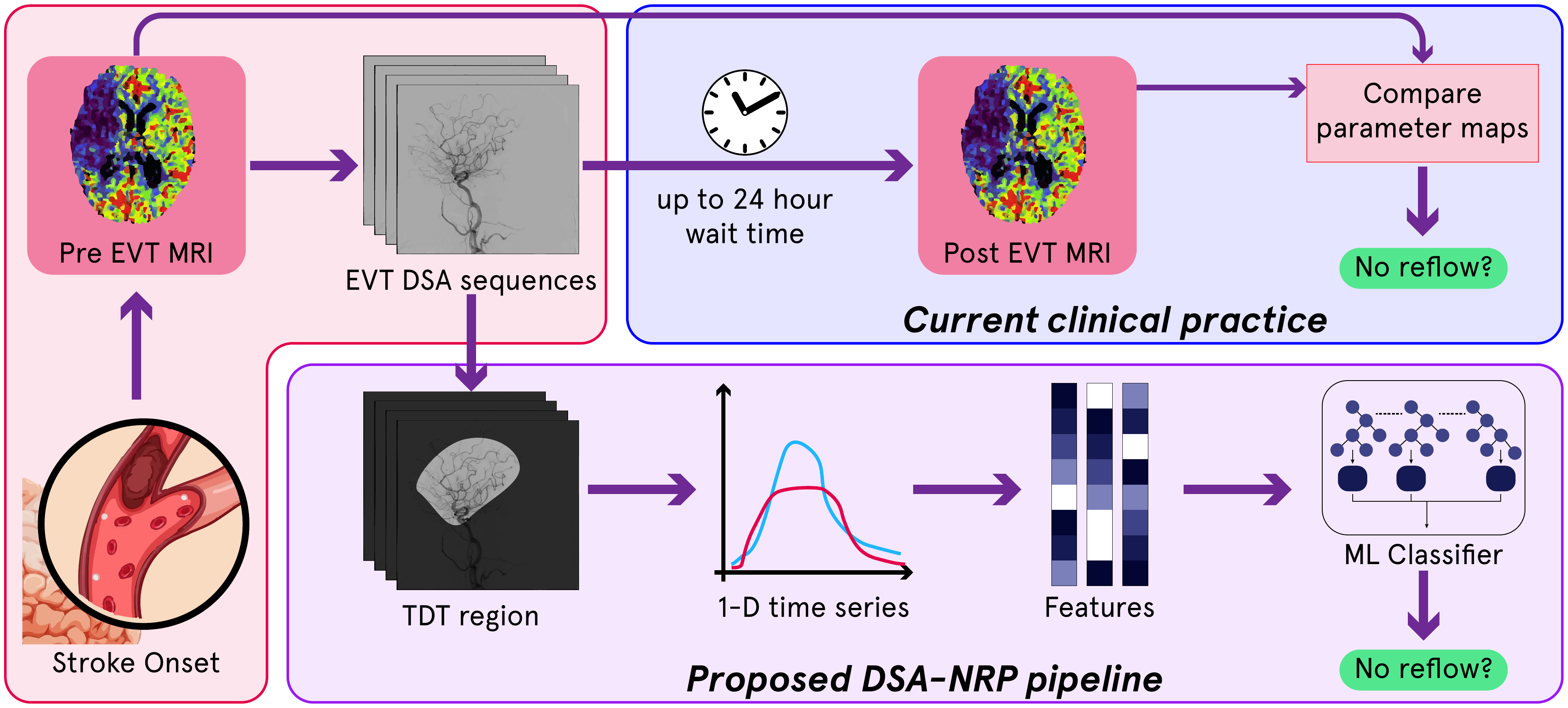}
    \caption{\textbf{Overview of clinical workflow and the proposed DSA-NRP pipeline for no-reflow prediction.} The proposed approach enables immediate, intra-procedural prediction of no-reflow using features extracted from DSA sequences and machine learning, reducing reliance on delayed follow-up MRI and supporting earlier, more targeted intervention.}
    \label{fig:intro}
\end{figure*}

Upon admission, stroke patients receive urgent neuroimaging evaluation, typically non-contrast computed tomography (CT)/magnetic resonance imaging (MRI), along with CT/MR perfusion and angiography of the head and neck to confirm large-vessel occlusion and assess the infarct core and salvageable tissue~(\autoref{fig:intro} (\textit{left})). Eligible patients undergo EVT, guided by digital subtraction angiography (DSA), and recanalization success is graded using the modified Thrombolysis in Cerebral Infarction (mTICI) score~\cite{patel_hyperacute_2020}. The mTICI scale ranges from 0 (no perfusion) to 3 (complete perfusion), with 2c or higher considered successful macrovascular recanalization. However, mTICI 2c or 3 does not ensure restored microvascular perfusion. The \emph{no-reflow} phenomenon, characterized by persistent tissue hypoperfusion despite macrovascular recanalization, is well documented~\cite{mujanovic_no-reflow_2024, horie_recanalization_2025, schiphorst_tissue_2021}. Currently, no-reflow is identified only through follow-up perfusion studies several hours after thrombectomy, which limits opportunities for immediate intervention~\cite{ng_prevalence_2022, liebeskind_etici_2019, mujanovic_association_2023}.

No-reflow is particularly relevant for proximal large-vessel occlusions involving the internal carotid artery (ICA) and the M1 segment of the middle cerebral artery (MCA), as these are the primary EVT targets with robust support from randomized trials like SWIFT PRIME, REVASCAT, and ESCAPE~\cite{powers_2015_2015, cho_mechanical_2021}. Conversely, recent randomized clinical trials showed no significant EVT benefit for distal medium-vessel (M2) occlusions, reducing the clinical relevance of no-reflow in these cases~\cite{liyis_mechanical_2023,goyal_endovascular_2025}. Accurate detection of no-reflow relies on perfusion imaging sensitive to subtle microvascular hemodynamic disturbances, which in this study are characterized by examining relative cerebral blood volume and flow (rCBV/rCBF) within the infarct core, as derived from MR or CT perfusion imaging~\cite{mujanovic_no-reflow_2024, kan_no-reflow_2023, abdalkader_neuroimaging_2023}. MRI, particularly within the first three hours post-onset, provides superior sensitivity to ischemia compared to CT: DWI, FLAIR, and GRE/SWI sequences effectively capture cytotoxic edema, vasogenic edema, and hemorrhagic transformation~\cite{arnould_comparison_2004, kan_no-reflow_2023, hwang_comparative_2012}.

To address the limitations of delayed MRI assessment, we propose \emph{DSA-NRP}, a pilot machine learning framework that predicts no-reflow immediately after EVT by integrating clinical features with intra-procedural DSA sequences (\autoref{fig:intro}). Early identification of patients at risk enables timely interventions such as neurocritical care~\cite{sheriff_dynamic_2020, migdady_current_2023}, individualized blood pressure management~\cite{peng_blood_2021, dong_blood_2024}, thrombus aspiration~\cite{svilaas_thrombus_2008}, or intra-arterial thrombolysis~\cite{kaesmacher_safety_2020, renu_effect_2022}. We hypothesize that DSA sequences obtained during EVT contain perfusion patterns in the target downstream territory (TDT)~\cite{su_autotici_2021} that are indicative of microcirculatory status, and these patterns can be extracted to predict no-reflow before follow-up imaging. Our contributions:
\begin{enumerate}[left=0pt, align=left]
    \item We present the first ML framework to predict post-thrombectomy no-reflow using structured, interpretable features derived from intra-procedural DSA sequences.
    
    \item We demonstrate, in this preliminary cohort, that DSA-derived features: specifically, statistical features and time-series signal descriptors, used with simple, interpretable classifiers outperform traditional clinical baselines, enabling early identification of patients at risk of no-reflow immediately following EVT.
    
    \item We conduct a detailed feature importance analysis, revealing physiologically meaningful predictors such as peak tracer intensity, temporal variance, and flow decay, which are associated with downstream perfusion failure. These results offer novel insights into DSA-based hemodynamic signatures of microvascular dysfunction.
\end{enumerate}

\section{Related Work}

\subsection{Predicting No-Reflow after EVT in AIS}
Early prediction of no-reflow is vital for managing AIS. Clinical factors such as age, hypertension, diabetes, high NIHSS, and large infarct core are strongly linked to no-reflow and futile recanalization~\cite{huang_comprehensive_2024}. Imaging biomarkers, like the modified Capillary Index Score (mCIS), quantify delayed microvascular filling and  correlate with neurological decline and hemorrhagic complications~\cite{nicolini_no-reflow_2023}. MRI and CT perfusion imaging within 24 hours post-procedure reveal persistent hypoperfusion, predicting worse outcomes~\cite{ng_prevalence_2022}. Recent reviews highlight advanced modalities (arterial spin labeling MRI, transcranial Doppler, AI-driven qTICI) for objective no-reflow quantification~\cite{kan_no-reflow_2023, prasetya_qtici_2021}. Our study builds on these multimodal indicators, uniquely combining intraprocedural DSA videos and clinical variables to extract angiographic features for accurate no-reflow prediction~\cite{sun_predictors_2023}.

\subsection{DSA Biomarkers for Stroke Intervention Outcomes}
DSA sequences acquired during thrombectomy provide biomarkers for stroke outcome assessment, including recanalization quality, microvascular perfusion, and residual thrombus. Angiographic indices like the modified Capillary Index Score (mCIS) quantify delayed or absent capillary filling despite vessel reopening, identifying microvascular no-reflow and correlating with early neurological decline and hemorrhagic transformation~\cite{nicolini_no-reflow_2023}. Automated and semi-automated techniques such as quantitative TICI (qTICI) use biplane DSA processing to objectively measure recanalization, achieving predictive accuracy comparable to experts~\cite{prasetya_qtici_2021, sabieleish_image_2021}. Deep learning models analyzing DSA temporal flow patterns reliably detect residual thrombi, allowing real-time incomplete recanalization identification~\cite{mittmann_deep_2022}. Other biomarkers, such as parenchymal blush and collateral filling, also predict complications~\cite{kan_no-reflow_2023}. Our work builds on these advances, using quantitative temporal DSA features for binary no-reflow classification.

\subsection{ML and DL in Stroke Intervention Analysis}
ML and DL methods are increasingly used to improve outcome prediction and detect complications in stroke interventions. Radiomics-based ML approaches leveraging pretreatment MRI can predict recanalization outcomes (mTICI 2c or 3) via infarct core, collateral status, and clot features, achieving around 75\% accuracy~\cite{zhang_machine_2021}. DL models using baseline CT and MRI anticipate the first-pass effect and difficult recanalization~\cite{zhang_deep_2024}, while radiomics models trained on paired pre- and 24-hour post-procedure NCCT scans predict 90-day outcomes but cannot inform intraprocedural decisions~\cite{da_ros_ensemble_2024}. In contrast, real-time analyses with 3D CNNs and recurrent DL models automate DSA-based classification of recanalization efficacy and microvascular flow~\cite{kelly_deep_2023, mittmann_deep_2022}. Cross-view fusion networks further improve automated TICI grading~\cite{xu_cvfsnet_2025}. Our approach, using interpretable ML on DSA and clinical features, optimizes predictive performance and clinical applicability~\cite{zhang_machine_2021, zhang_deep_2024, da_ros_ensemble_2024, kelly_deep_2023}.

\begin{table}
\centering
\caption{Demographic comparisons between patients with reflow and no-reflow status, as determined by~\cite{ng_prevalence_2022}. IQR: interquartile range; NIHSS: national institutes of health stroke scale; ICA: internal cerebral artery; TTR: time to recanalization (hours).}
\label{tab:demographics}
\renewcommand{\arraystretch}{1.2}
\begin{tabular*}{0.7\columnwidth}{@{\extracolsep{\fill}}|l@{\hspace{2em}}|r|r|@{\extracolsep{\fill}}}
\hline
\textbf{Variable} & \textbf{Reflow} $(n=32)$ & \textbf{No reflow} $(n=7)$ \\
\hline
Sex = Male (\%) & 9 (28.1) & 2 (28.6) \\
Hemisphere = Right (\%) & 10 (31.2) & 4 (57.1) \\
Age (median [IQR]) & 80 [64, 86] & 69 [58, 77] \\
NIHSS (median [IQR]) & 15 [11, 20] & 18 [9, 20] \\
TTR (median [IQR]) & 5.9 [3.5, 13.1] & 4.1 [4.0, 14.4] \\
\hline
\textbf{EVT mTICI scores} & & \\
2c (\%) & 14 (43.8) &  5 (71.4) \\
3  (\%) & 18 (56.2) &  2 (28.6) \\
\hline
\textbf{Occlusion location} & & \\
M1 (\%)     &  6 (18.8) &  2 (28.6) \\
ICA (\%)    &  5 (15.6) &  0 (\phantom{1}0.0)  \\
M1/ICA (\%) & 21 (65.6) &  5 (71.4) \\
\hline
\textbf{Race} & & \\
White/Caucasian (\%) & 17 (53.1) & 3 (42.9) \\
Black (\%)           &  6 (18.8) & 1 (14.3) \\
Other (\%)           &  9 (28.1) & 3 (42.9) \\
\hline
\textbf{Number of passes} & & \\
$n = 1$ (\%)     & 18 (56.2) &  3 (42.9) \\
$n = 2$ (\%)     &  7 (21.9) &  2 (28.6) \\
$n \geq 3$ (\%)  &  7 (21.9) &  2 (28.6) \\
\hline
\textbf{Comorbidities} & & \\
Hypertension (\%)   & 16 (50.0) &  3 (42.9) \\
Diabetes (\%)       &  7 (21.9) &  3 (42.9) \\
Atrial Fib (\%)     & 11 (34.4) &  2 (28.6) \\
Hyperlipidemia (\%) &  9 (28.1) &  2 (28.6) \\
Coag. disorder (\%) &  8 (25.0) &  0 (\phantom{1}0.0) \\
\hline
\end{tabular*}
\end{table}

\section{Methods}

\subsection{Dataset}

The dataset comprises a retrospective cohort of 39 patients (\autoref{tab:demographics}) who experienced large vessel occlusion AIS and underwent EVT at UCLA Medical Center between 2011 and 2024, with approval from the UCLA institutional review board (IRB\#18-0329). Inclusion criteria for this study were:  

\begin{enumerate}[left=0pt, align=left]
\item \emph{Patients with thrombus involving the M1 segment of the middle cerebral artery (MCA), the internal carotid artery (ICA), or combined M1/ICA branches}: This selection follows established guidelines and trial evidence demonstrating EVT efficacy for proximal large-vessel occlusions such as ICA and M1~\cite{powers_2015_2015, cho_mechanical_2021}, thereby ensuring a homogeneous study cohort with a well-defined therapeutic indication~\cite{liyis_mechanical_2023}.

\item \emph{Patients achieving mTICI scores of 2c or 3, consistent with successful recanalization}: Limiting the analysis to mTICI 2c or 3 isolates patients with adequate macrovascular recanalization, allowing true microvascular no-reflow phenomena to be evaluated, in line with prior studies in this field~\cite{mujanovic_no-reflow_2024, horie_recanalization_2025, schiphorst_tissue_2021, ng_prevalence_2022, liebeskind_etici_2019, mujanovic_association_2023}.

\item \emph{Patients with both pre-procedure and post-procedure MRI  available}: MRI offers higher sensitivity for ischemia detection and enables reliable perfusion mapping for assessing no-reflow~\cite{abdalkader_neuroimaging_2023, hwang_comparative_2012, kan_no-reflow_2023}, consistent with our institutional MR-first protocols and reducing variability related to imaging modality~\cite{arnould_comparison_2004}.

\end{enumerate}

From an initial registry of $(n=710)$ LVO AIS patients presenting between 2011 and 2024, we identified eligible subjects by sequentially selecting for EVT treatment $(n=638)$, M1/ICA occlusion location $(n=422)$, availability of paired pre- and post-procedure MRI $(n=167)$, availability of paired pre- and post-procedure MR perfusion imaging $(n=87)$, and successful recanalization defined as mTICI 2c or 3 $(n=39)$. Patients with no-reflow did not differ significantly from those with reflow with respect to baseline demographics or clinical severity. Age demonstrated a moderate effect size but did not reach statistical significance (Mann–Whitney U, $p=0.096$), while baseline NIHSS and time to recanalization showed no evidence of group differences. Sex, race, hemisphere, occlusion location, number of device passes, and final EVT mTICI scores were also comparable between groups (all $p>0.20$). Similarly, no statistically significant associations were observed between no-reflow and common vascular comorbidities, including hypertension, diabetes mellitus, atrial fibrillation, hyperlipidemia, or coagulation disorders (Fisher’s exact test, all $p>0.30$). Overall, baseline demographic, clinical, and procedural characteristics appeared broadly balanced between groups, and none of the examined variables demonstrated a statistically detectable association with no-reflow in this cohort, likely reflecting limited statistical power.

Each patient had a series of DSA sequences captured during EVT, which were manually annotated by an expert interventional radiologist using a custom GUI\footnote{\url{https://github.com/cao826/dsa-annotations}}. The annotations identified the specific pre- and post-procedure DSA series used to locate the thrombus and determine the mTICI score. The physician marked the thrombus on one frame, and this location was propagated across all frames, creating image patch sequences. For each patient, four DSA sequences were selected: paired pre- and post-procedure series in both anteroposterior (AP) and lateral views.

To determine no-reflow status, we adopted the quantitative definition established by Ng \textit{et al.}, chosen for its rigorous use of perfusion asymmetry in confirmed infarct zones~\cite{ng_prevalence_2022}. Following their protocol, no-reflow was defined as a $>15\%$ reduction in median relative cerebral blood volume (rCBV) or blood flow (rCBF) within the infarct core compared to the contralateral hemisphere. To automate this process, we generated infarct Regions of Interest (ROIs) from follow-up Diffusion-Weighted Imaging (DWI) using a U-Net architecture. This model was pre-trained on the BRATS21 dataset~\cite{baid2021rsna} and fine-tuned on the ISLES 2022 dataset using the top-ranking `SEALS' methodology~\cite{de2024robust}. The resulting ROI was mirrored across the longitudinal fissure to establish the contralateral baseline, with areas of hemorrhage automatically excluded prior to sampling. For full implementation details regarding this automated pipeline and label generation, we refer readers to our prior work~\cite{OlivaresReboredo2025ExploringAssessment}.

\subsection{DSA pre-processing}
Each DSA sequence is a grayscale video (1024~$\times$~1024 pixels, $\sim$6 seconds, 3 fps) visualizing iodinated contrast flow through cerebral vessels during EVT. Analyzing both pre- and post-procedure sequences is vital for detecting blood flow changes after thrombus removal and identifying no-reflow. Accurately isolating the TDT, which is the region affected by occlusion enables assessment of local perfusion changes. However, consistent TDT analysis is challenging due to variability in DSA acquisition: differences in image zoom, head orientation, and operator focus complicate comparisons across sequences and patients, making standardized, reliable analysis crucial for meaningful perfusion evaluation.

To standardize analysis and address variability, we implemented structured preprocessing. The TDT region, which represents the cerebral territory supplied by the occluded artery, was delineated by manually annotating the pixel-wise minimum intensity projection (\emph{minProjection}) image, following prior methods~\cite{prasetya_qtici_2021, su_autotici_2021}. Although cortical coverage was sometimes limited due to patient motion artifacts and the frequent use of zoomed-in post-procedure imaging, which reduced the field of view, this occurred randomly and thus did not systematically bias the comparisons between reflow and no-reflow patients. The \emph{minProjection} image summarizes regions traversed by contrast during the sequence. Two graduate students performed the annotations under expert interventional radiologist supervision. To ensure spatial alignment, pre- and post-procedure DSA sequences were manually registered using minimal affine transformations and anatomical landmarks, aligning both DSA and \emph{minProjection} images. This process enabled consistent TDT annotation across sequences. The \emph{minProjection} image was used only for TDT localization, while all further analyses used the full frame-by-frame DSA data to preserve temporal perfusion dynamics.

\begin{figure}[!tbp]
    \centering
    \includegraphics[width=0.7\columnwidth]{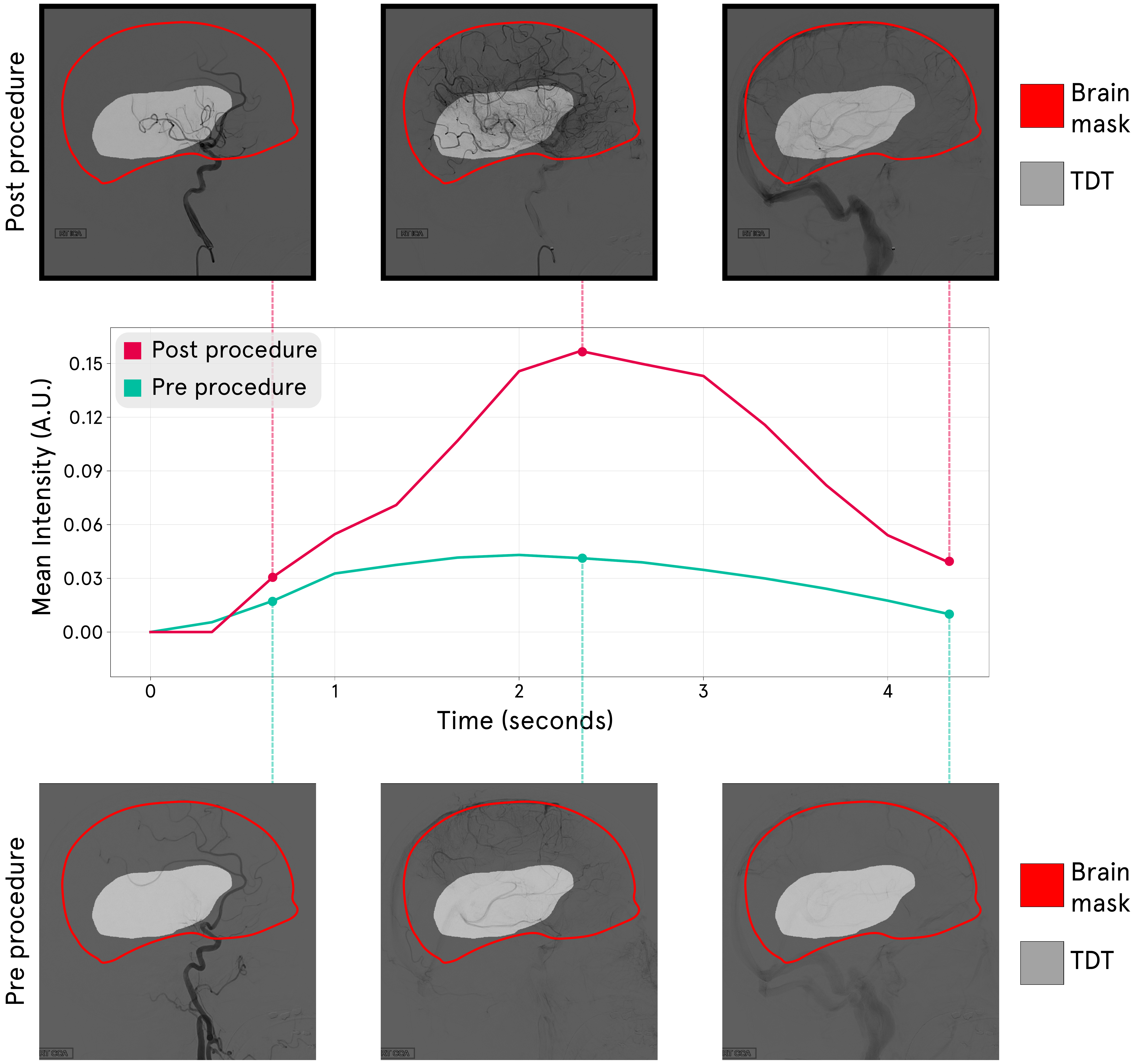}
    \caption{\textbf{Conversion of DSA video data from the TDT region into a 1D time-series signal.} Mapping the mean pixel intensity within the TDT across all frames produces a compact time-series representation, enabling quantitative analysis of pre- and post-procedure perfusion dynamics for no-reflow prediction.}
    \label{fig:DSAtoTimeSeries}
\end{figure}

Subsequently, the binary TDT mask was uniformly applied to each frame of the DSA sequence, and mean pixel intensity within the TDT was computed, converting 3D spatio-temporal data into a 1D time-series signal. Lower mean intensities indicate greater tracer presence, robustly capturing perfusion dynamics~(\autoref{fig:DSAtoTimeSeries}, top row). Prior to extraction, each DSA sequence underwent grayscale intensity inversion that ensured brighter pixels corresponded to higher tracer concentrations, resulting in intuitive unimodal perfusion curves: intensity rises in arterial/capillary phases, peaks, then declines during venous outflow~(\autoref{fig:DSAtoTimeSeries}). This facilitates clear interpretation and patient-to-patient comparisons. Temporal misalignment due to operator-dependent recording and injection timing was addressed by synchronizing signals: the onset of contrast was identified as the first frame where the absolute first derivative exceeded 0.01. Signals were truncated from this point for consistent alignment.

\begin{figure*}[!htbp]
    \centering
    \includegraphics[width=\textwidth]{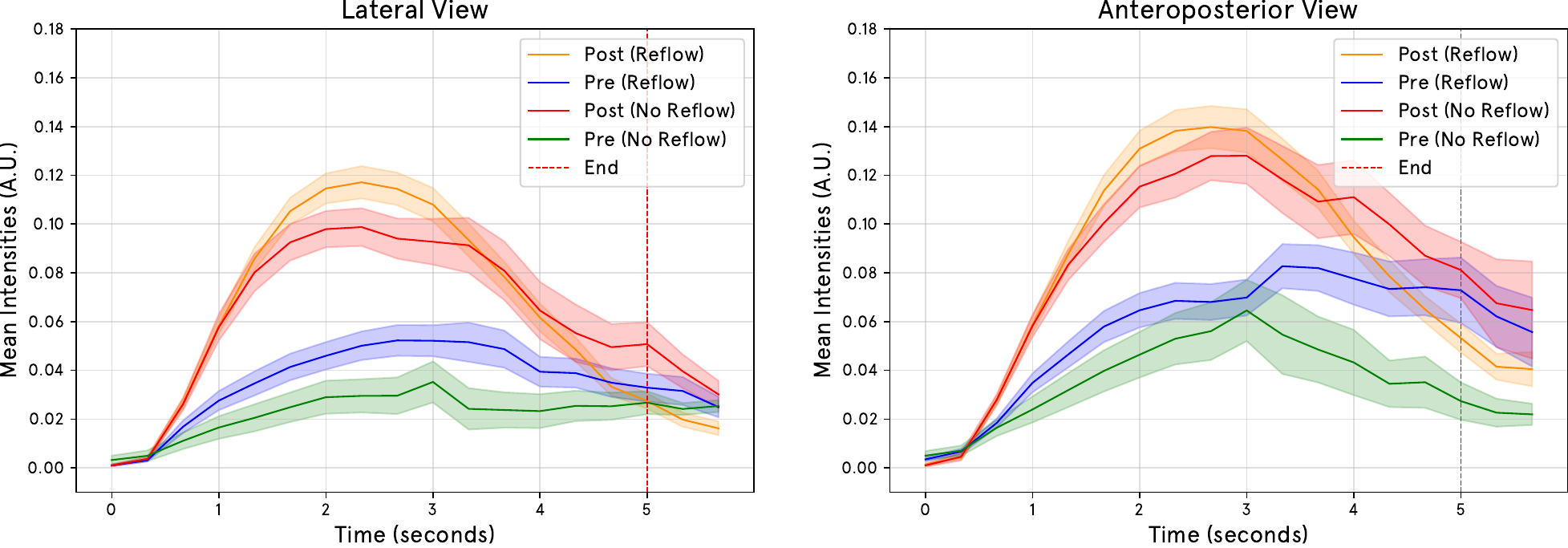}
    \caption{\textbf{Mean intensity time-series curves from the TDT region, averaged across patients for each outcome group.} Distinct temporal profiles between reflow and no-reflow cases are observed in both lateral and AP views, highlighting key differences in post-procedural perfusion dynamics. In the analysis, signals are truncated at $t=5$ seconds.}
    \label{fig:mean_curves}
\end{figure*}

\subsection{Feature extraction}

Prior studies~\cite{jann_implication_2016,lee_automatic_2017,lu_digital_2021} indicate that the most clinically relevant tracer dynamics: the arterial and capillary phases, occur within the first five seconds after contrast enters the TDT~(\autoref{fig:mean_curves}), so feature extraction was limited to this window. Features were derived from pre- and post-procedure 1D intensity signals, $x_{\text{pre}}(t)$ and $x_{\text{post}}(t)$, from AP and lateral DSA sequences ($t$: frame index). As a baseline, we used clinical features \texttt{(CLN)}: age, gender, and final mTICI, reflecting standard risk stratification practices and providing a strong reference for assessing the value of DSA-derived features. Motivated by prior work~\cite{Sano2025PredictorsCirculation.,Kiani2025NeuroimagingMeta-Analysis.}, we initially considered an expanded set of clinical variables beyond age, sex, and mTICI score, including NIHSS, time to recanalization, number of passes, and common vascular comorbidities; however, as this expansion did not yield additional predictive benefit, we retained the more parsimonious subset for the clinical baseline and subsequent ablation analyses.

\subsubsection{Peak Dynamics Analysis~\texttt{(PEAK)}}
These features capture dominant perfusion peaks in pre- and post-procedure signals, reflecting arterial/capillary phases. Their shape, and amplitude provide insights into vascular integrity and recanalization efficacy.

\paragraph{\textbf{Difference in peak height} \texttt{(peakHeight)}}
\begin{equation}
\Delta H_{\text{peak}} = \max(x_{\text{post}}(t)) - \max(x_{\text{pre}}(t))
\end{equation}
Captures the change in maximum tracer intensity, serving as a proxy for net contrast delivery into the region of interest.

\paragraph{\textbf{Difference in peak width at half maximum} \texttt{(peakWidth)}}
Given the perfusion signal is largely unimodal, we define $\Delta t_{\text{pre}}$ as the duration between the first and last time points where the pre-procedure signal remains above half of its maximum value:
\begin{align}
\Delta t_{\text{pre}} = &\max\{ t \mid x_{\text{pre}} > 0.5 \cdot \max(x_{\text{pre}})\} \notag\\ 
- &\min\{ t \mid x_{\text{pre}} > 0.5 \cdot \max(x_{\text{pre}}) \}
\end{align}
An analogous definition applies for $\Delta t_{\text{post}}$, and the final feature is computed as:
\begin{equation}
\Delta W_{\text{peak}} = \Delta t_{\text{post}} - \Delta t_{\text{pre}}
\end{equation}
It captures how long contrast intensity remains elevated post-recanalization relative to baseline.

\paragraph{\textbf{Peak intensity ratio} \texttt{(peakRatio)}}
\begin{equation}
R_{\text{peak}} = \max(x_{\text{post}}(t)) / \max(x_{\text{pre}}(t))
\end{equation}
Quantifies the relative gain in peak perfusion; values less than one may suggest suboptimal reperfusion.

\paragraph{\textbf{Difference in peak slope} \texttt{(peakSlope)}}
\begin{equation}
\Delta m_{\max} = \max\left(\nabla x_{\text{post}}(t)\right) - \max\left(\nabla x_{\text{pre}}(t)\right)
\end{equation}
Measures the sharpest rate of contrast inflow. Overall, early temporal peak features may reflect upstream hemodynamic effects associated with microvascular dysfunction and no-reflow~\cite{Qiu2023ARecanalization,Cipolla2014PostischemicArterioles}.

\subsubsection{Statistical Intensity Profiling~\texttt{(SIPS)}}
These features capture statistical properties: mean, variability, skewness, and kurtosis of pre- and post-procedure signals, reflecting global changes in perfusion quality and tracer delivery. We denote the means and standard deviations as $\mu_{\text{pre}}, \sigma_{\text{pre}}, \mu_{\text{post}}, \text{ and } \sigma_{\text{post}}$.

\paragraph{\textbf{Difference in mean intensity} \texttt{(meanIntensity)}} Reflects the overall change in signal intensity between pre- and post-procedure sequences, providing a coarse indicator of contrast uptake in the region.
\begin{equation}
\Delta \mu = \mu_{\text{post}} - \mu_{\text{pre}}    
\end{equation}

\paragraph{\textbf{Difference in intensity standard deviation} \texttt{(stdDevIntensity)}} Captures variability in perfusion, with larger standard deviations post-procedure potentially indicating turbulent or uneven flow.
\begin{equation}
\Delta \sigma = \sigma_{\text{post}} - \sigma_{\text{pre}}
\end{equation}

\paragraph{\textbf{Difference in minimum intensity} \texttt{(minIntensity)}} Highlights changes in baseline signal floor, which may be sensitive to low-flow or poorly perfused regions.
\begin{equation}
\Delta I_{\min} = \min(x_{\text{post}}(t)) - \min(x_{\text{pre}}(t)) 
\end{equation}

\paragraph{\textbf{Ratio of mean intensities} \texttt{(meanIntensityRatio)}} Normalizes intensity change to baseline perfusion, providing a relative measure of contrast delivery independent of scale.
\begin{equation}
R_{\mu} = \mu_{\text{post}} / \mu_{\text{pre}}
\end{equation}

\paragraph{\textbf{Difference in skewness} \texttt{(skewness)}} Measures asymmetry in the intensity distribution. Post-procedure increases in skewness may reflect lingering high-intensity voxels due to contrast stasis.
\begin{equation}
\Delta \text{Sk.} = \mathop{\mathbb{E}}\left[\left(\frac{x_{\text{post}}-\mu_{\text{post}}}{\sigma_{\text{post}}}\right)^3\right] - \mathop{\mathbb{E}}\left[\left(\frac{x_{\text{pre}}-\mu_{\text{pre}}}{\sigma_{\text{pre}}}\right)^3\right]
\end{equation}

\paragraph{\textbf{Difference in kurtosis} \texttt{(kurtosis)}} Assesses peakedness or tail heaviness in the signal distribution; higher post-procedure kurtosis may suggest heterogeneous or abnormal tracer retention.
\begin{equation}
\Delta \text{Kt.} = \mathop{\mathbb{E}}\left[\left(\frac{x_{\text{post}}-\mu_{\text{post}}}{\sigma_{\text{post}}}\right)^4\right] - \mathop{\mathbb{E}}\left[\left(\frac{x_{\text{pre}}-\mu_{\text{pre}}}{\sigma_{\text{pre}}}\right)^4\right]
\end{equation}

\subsubsection{Temporal Flow Comparison~\texttt{(FLOW)}}
These features quantify timing-related tracer perfusion dynamics, including onset, peak, and clearance speed, highlighting changes in filling and washout between pre- and post-procedure signals. We define $\mathbf{t}_{\alpha}(x)$ as the earliest frame reaching $\alpha \cdot \max(x(t)),\, \alpha\in [0,1]$.

\paragraph{\textbf{Difference in time to reach 50\% of peak intensity} \texttt{(timeTo50Max)}} Measures the timing shift in initial tracer inflow; delays may indicate impaired perfusion or collateral dependency.
\begin{equation}
\Delta t_{50\%} = \mathbf{t}_{0.5}(x_{\text{post}}) - \mathbf{t}_{0.5}(x_{\text{pre}})
\end{equation}

\paragraph{\textbf{Difference in time to reach peak intensity} \texttt{(timeToPeak)}} Captures changes in tracer arrival dynamics; altered timing post-procedure can reflect differences in flow velocity.
\begin{equation}
\Delta t_{\text{peak}} = \mathbf{t}_{1.0}(x_{\text{pre}}) - \mathbf{t}_{1.0}(x_{\text{post}})
\end{equation}

\paragraph{\textbf{Difference in decay time} \texttt{(decayTime)}} Let $t_{\text{decay}}(x)$ represent the frame index at which the signal $x$ first drops to 10\% of its maximum intensity after achieving its peak. 
\begin{equation}
t_{\text{decay}}(x) = \mathbf{t}_{0.9}(x), \quad \text{where } \mathbf{t}_{0.9}(x) > \mathbf{t}_{1.0}(x)
\end{equation}
Then the decay time difference is expressed as:
\begin{equation}
\Delta t_{\text{decay}} = t_{\text{decay}}(x_{\text{post}}) - t_{\text{decay}}(x_{\text{pre}})
\end{equation}
This evaluates how quickly tracer exits the region; prolonged retention may signal downstream microvascular stasis.

\paragraph{\textbf{Difference in plateau duration} \texttt{(plateauDuration)}} Quantifies how long the signal remains flat; longer plateaus post-procedure may reflect sluggish flow or incomplete clearance.
\begin{equation}
\Delta T_{\text{plateau}} = N_{\text{post}} - N_{\text{pre}}, \,\ N = \sum_{t} \left( |\nabla x(t)| < 0.01 \right)
\end{equation}

\paragraph{\textbf{Signal correlation between pre- and post- sequences} \texttt{(signalCorrelation)}} Measures overall signal similarity; low correlation indicates significant hemodynamic changes due to intervention.
\begin{equation}
\rho = \frac{\text{cov}(x_{\text{pre}}(t), x_{\text{post}}(t))}{\sigma_{\text{pre}}\sigma_{\text{post}}}
\end{equation}

Collectively, \texttt{PEAK}, \texttt{SIPS} and \texttt{FLOW} form a comprehensive set of \textit{Time-Series Features} (\texttt{TSF}) extracted from the 1D intensity signals.

\begin{figure*}[!tbp]
    \centering
    \includegraphics[width=\textwidth]{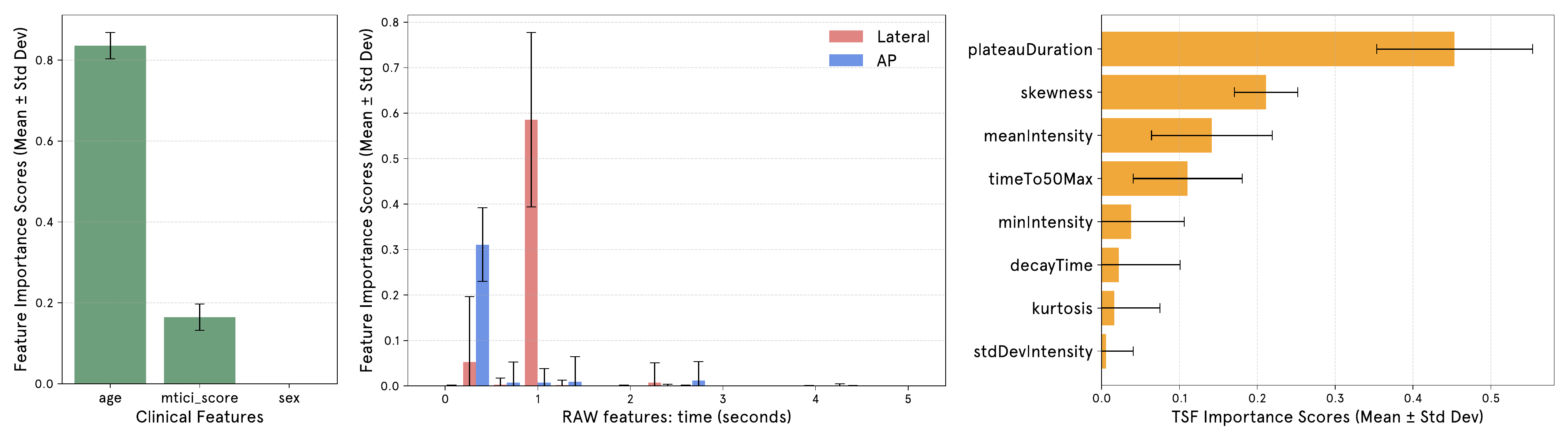}
    \caption{\textbf{Feature importance for no-reflow prediction for different feature sets.} \textit{(Left:)} In \texttt{CLN}, age was the most important feature. \textit{(Middle:)} Earlier time-points in \texttt{RAW} contributed strongly to model performance as compared to later ones. \textit{(Right:)} The combined \texttt{SIPS+FLOW} set extracted from lateral post-EVT signals were key contributors, underscoring the physiological relevance of intensity dynamics and temporal descriptors in predicting no-reflow.}
    \label{fig:results}
\end{figure*}

\subsubsection{Fixed-length 1D Time-Series Data Points}
To enable consistent comparison across patients, post-procedure DSA time-series data points are standardized into fixed-length feature vectors \texttt{(RAW)} of $L=15$ frames. For signals shorter than $L$, missing values are extrapolated using exponential decay:
\begin{equation}
f[t] = f[n] \cdot \frac{e^{-\beta(t-n+1)} - e^{-\beta(L-n)}}{1 - e^{-\beta(L-n)}}
\end{equation}
for $t = n+1$ to $L$. Here, $f[n]$ is the last observed value and $\beta$ controls decay steepness. This approach ensures a smooth transition to near-zero at the end, reflecting physiological tracer wash-out, and captures essential perfusion dynamics for no-reflow prediction.

\subsection{Univariate feature analysis}
To evaluate whether individual time-series features exhibited directional differences consistent with their physiological motivation, we performed hypothesis-driven univariate analyses across all 30 features derived from lateral and AP views. For each feature, we tested a one-sided Mann-Whitney U hypothesis reflecting the expected change in the no-reflow group. Overall, univariate statistical significance was limited, consistent with the small cohort size and the exploratory nature of these tests. For features computed from combined pre- and post-procedure signals, most lateral and AP features demonstrated directionally consistent but non-significant trends ($p > 0.05$). One exception was lateral decay time, which was significantly lower in the no-reflow group ($p = 0.021$), supporting altered post-peak washout dynamics. In the AP view, minimum intensity was significantly increased in no-reflow patients ($p = 0.013$), consistent with impaired microvascular clearance. When restricting analysis to post-procedure-only features, lateral decay time again showed a significant reduction in the no-reflow group ($p = 0.026$). Several additional lateral features exhibited borderline trends in the hypothesized direction, including peak slope, skewness, and plateau duration, whereas AP-view features remained largely non-significant. Collectively, these findings suggest that while individual features are underpowered to demonstrate strong univariate effects, they exhibit biologically plausible directional behavior that contributes meaningfully when integrated within a multivariate learning framework.

\subsection{ML frameworks and training implementation}
\label{ss:implementation}
Given the limited cohort size, we adopted a leave-one-out cross-validation (LOOCV) strategy to evaluate model performance while minimizing overfitting. In each iteration, a single patient was held out as the test case, and the model was trained on the remaining subjects. This procedure was repeated until every patient had served as the held-out test case exactly once, yielding a single out-of-sample prediction per subject. Performance metrics were then computed using the aggregated predictions across all LOOCV iterations, providing an unbiased estimate of model generalization under maximal data utilization. Feature categories were preprocessed separately: \texttt{CLN} and \texttt{TSF} were encoded and normalized using only training-fold statistics to prevent data leakage, and no feature selection was performed. \texttt{RAW} features were used as-is to preserve temporal structure. We conducted ablation studies by training models on individual feature sets, all pairwise combinations, and the full set. Multiple classifiers were evaluated, including random forest, support vector classifier, multi-layer perceptron, and gradient boosting classifier. The gradient boosting classifier, with 10 estimators, max depth of 3, and a 0.9 learning rate, was selected for its strong performance and stability. Models using \texttt{CLN} alone served as the baseline. Improvements over baseline were assessed for significance with a one-sided Wilcoxon signed-rank test. Secondary comparisons against \texttt{CLN} were adjusted for multiple testing using the Holm–Bonferroni procedure, while the primary comparison between \texttt{CLN} and the best-performing model was pre-specified and evaluated without correction. The source code and implementation details are available here\footnote{\url{https://github.com/ShreeramAthreya/DSA_NRP}}.

\section{Results}
This retrospective pilot study included 39 EVT-treated AIS patients (2011–2024, UCLA IRB\#18-0329) with M1/ICA occlusions, successful recanalization (mTICI 2c–3), and paired pre- and post-procedure MRI scans. Models were trained using clinical features (\texttt{CLN}) and DSA-derived features: frame-level 1D time-series signals (\texttt{RAW}) and statistical time-series features (\texttt{TSF}) extracted from lateral and AP projections, as well as ensemble combinations of both views. The clinical baseline achieved an AUROC of 0.7768 with an accuracy of 0.7949, indicating modest discriminative ability (\autoref{tab:model_performance}). In this preliminary cohort, DSA-derived features were associated with improved predictive performance compared to the clinical baseline. While \texttt{CLN} demonstrated reasonable specificity (0.8125), its precision (0.4545) and F1 score (0.5556) were limited, reflecting reduced reliability for identifying no-reflow cases in isolation. In contrast, DSA-derived features substantially improved predictive performance. Models trained using \texttt{RAW} features achieved an AUROC of 0.9152 ($p = 0.013$) and accuracy of 0.9231, with perfect recall and strong specificity (0.9063). Similarly, the \texttt{TSF} configuration achieved an AUROC of 0.9018 ($p = 0.009$) with high accuracy (0.8974) and perfect recall when using lateral-view features alone. The highest overall performance was observed with the combined \texttt{CLN+TSF+RAW} ensemble model (AUROC 0.9330, $p = 0.006$; accuracy 0.9231; precision 0.7000), although this improvement was not statistically different from models using DSA-derived features alone. Notably, augmenting \texttt{RAW} or \texttt{TSF} models with clinical features (\texttt{CLN+RAW}, \texttt{CLN+TSF}) did not yield consistent performance gains, suggesting that baseline clinical variables contribute limited additional information beyond DSA-derived temporal features. These pilot findings suggest a prominent role for DSA-based signals in no-reflow prediction, with limited incremental contribution from clinical features, although the results should be interpreted cautiously given the modest sample size.

\begin{table*}[!tbp]
\centering
\caption{Model performance across different feature configurations. \texttt{CLN}: Clinical baseline, \texttt{RAW}: 1D Data points, \texttt{TSF}: Time-series features. * Indicates Holm–Bonferroni adjusted p-values.}
\renewcommand{\arraystretch}{1.5}
\begin{tabular*}{\textwidth}{@{\extracolsep{\fill}}|l|l|*{6}{r|}@{\extracolsep{\fill}}}
\hline
\textbf{Feature Set} & \textbf{AUROC}  &\textbf{Accuracy}& \textbf{Recall} &\textbf{Specificity}& \textbf{F1 Score} & \textbf{Precision} & \textbf{View} \\
\hline
\texttt{CLN} (baseline)         & 0.7768          & 0.7949          & 0.7143          & 0.8125          & 0.5556          & 0.4545          & - \\
\hline
\texttt{RAW}         & 0.9152 ($p=0.013$)*          & \textbf{0.9231} & \textbf{1.0000} & \textbf{0.9063} & \textbf{0.8235} & \textbf{0.7000} & Ensemble\\
\hline
\texttt{TSF}         & 0.9018 ($p=0.009$)*         & 0.8974          & \textbf{1.0000} & 0.8750          & 0.7778          & 0.6364          & Lateral \\
\hline
\hline
\texttt{CLN+RAW}     & 0.8571 ($p=0.018$)*         & 0.8205          & \textbf{1.0000} & 0.7813          & 0.6667          & 0.5000          & Lateral\\
\hline
\texttt{CLN+TSF}     & 0.8929 ($p=0.009$)*          & 0.8718          & \textbf{1.0000} & 0.8438          & 0.7368          & 0.5833          & Lateral \\
\hline
\texttt{TSF+RAW}     & 0.9152 ($p=0.018$)*          & \textbf{0.9231} & \textbf{1.0000} & \textbf{0.9063} & \textbf{0.8235} & \textbf{0.7000} & Ensemble\\
\hline
\hline
\texttt{CLN+TSF+RAW} & \textbf{0.9330} ($p=0.006$) & \textbf{0.9231} & \textbf{1.0000} & \textbf{0.9063} & \textbf{0.8235} & \textbf{0.7000} & Ensemble\\
\hline
\end{tabular*}
\label{tab:model_performance}
\end{table*}

\begin{table*}[!tbp]
\centering
\caption{Subgroup analyses for the best performing configuration. $n$: number of patients in the subgroup. \texttt{CLN}: Clinical baseline, \texttt{RAW}: 1D Data points, \texttt{TSF}: Time-series features. Median NIHSS was 15, and median age was 78 for our cohort.}
\renewcommand{\arraystretch}{1.5}
\begin{tabular*}{\textwidth}{@{\extracolsep{\fill}}|l|c|*{6}{c|}@{\extracolsep{\fill}}}
\hline
\textbf{Subgroup} & $n$ & \textbf{AUROC} & \textbf{Accuracy} & \textbf{Recall} & \textbf{Specificity} & \textbf{F1 Score} & \textbf{Precision} \\
\hline
\texttt{CLN+TSF+RAW} & 39 & \textbf{0.9330} &\textbf{ 0.9231} & \textbf{1.0000} & \textbf{0.9063} & \textbf{0.8235} & \textbf{0.7000} \\
\hline
\hline
$mTICI = 2c$ & 19 & 0.9571 & 0.9474 & 1.0000 & 0.9286 & 0.9091 & 0.8333 \\
\hline
$mTICI = 3$ & 20 & 0.8889 & 0.9000 & 1.0000 & 0.8889 & 0.6667 & 0.5000 \\
\hline
\hline
Sex $=$ Female & 28 & 0.9304 & 0.9286 & 1.0000 & 0.9130 & 0.8333 & 0.7143 \\
\hline
Sex $=$ Male   & 11 & 0.9444 & 0.9091 & 1.0000 & 0.8889 & 0.8000 & 0.6667\\
\hline
\hline
Race $=$ White & 20 & 0.9216 & 0.9000 & 1.0000 & 0.8824 & 0.7500 & 0.6000 \\
\hline
Race $=$ Other & 19 & 0.9500 & 0.9474 & 1.0000 & 0.9333 & 0.8889 & 0.8000 \\
\hline
\hline
Age $>$ 78~\cite{Messe2016WhyTPA} & 19 & 0.9706 & 0.9474 & 1.0000 & 0.9412 & 0.8000 & 0.6667 \\
\hline
Age $\leq$ 78~\cite{Messe2016WhyTPA} & 20 & 0.8933 & 0.9000 & 1.0000 & 0.8667 & 0.8333 & 0.7143 \\
\hline
\hline
Pre NIHSS $>$ 15 & 17 & 0.9615 & 0.9412 & 1.0000 & 0.9231 & 0.8889 & 0.8000 \\
\hline
Pre NIHSS $\leq$ 15 & 22 & 0.8947 & 0.9091 & 1.0000 & 0.8947 & 0.7500 & 0.6000 \\
\hline
\hline
num passes $=$ 1 & 21 & 0.9259 & 0.9048 & 1.0000 & 0.8889 & 0.7500 & 0.6000 \\ 
\hline
num passes $>$ 1 & 18 & 0.9464 & 0.9444 & 1.0000 & 0.9286 & 0.8889 & 0.8000 \\
\hline
\end{tabular*}
\label{tab:sub_analysis}
\end{table*}

\begin{table}[!tbp]
    \centering
    \caption{AUROC scores for time-series feature category ablations. The best-performing feature set in each combination is highlighted. * denotes the selected TSF configuration used for all subsequent analyses.}
    \renewcommand{\arraystretch}{1.5}
    \begin{tabular*}{\columnwidth}{@{\extracolsep{\fill}}|l|*{3}{r|}@{\extracolsep{\fill}}}
        \hline
        \textbf{Feature set} & \textbf{Combination} & \textbf{Post Only} & \textbf{Pre Only} \\ \hline
        \texttt{PEAK}        & 0.7321          & 0.7589           & 0.6607          \\
        \texttt{SIPS}        & 0.8125          & 0.7857           & 0.8080          \\
        \texttt{FLOW}        & \textbf{0.8616} & 0.7321           & 0.7679          \\ \hline
        \texttt{PEAK + SIPS} & 0.8438          & 0.6920           & 0.7545          \\ 
        \texttt{PEAK + FLOW} & 0.8482          & 0.8080           & 0.6920          \\
        \texttt{SIPS + FLOW} & 0.8393          & \textbf{0.9018*} & \textbf{0.8348} \\ \hline
        \texttt{PEAK + SIPS + FLOW} & 0.7857          & 0.8795           & 0.7679          \\ \hline
    \end{tabular*}
    \label{tab:ablations}
\end{table}

\subsection{Feature Importance and Physiological Relevance}

Feature importance analysis provides insight into the physiological drivers of no-reflow prediction across clinical, raw time-series, and statistical feature representations (\autoref{fig:results}). Among  \texttt{CLN}, age emerged as the dominant contributor, while sex and mTICI score had minimal influence, consistent with our ablation results indicating limited standalone predictive value of expanded clinical covariates. For models trained on post-procedure signals (\texttt{RAW}), the most influential features were consistently located in the early arterial phase of the contrast bolus, particularly in the lateral view. This pattern suggests that frame-level arterial-phase dynamics encode physiologically relevant information beyond summary statistics alone. Prior ischemia-reperfusion studies indicate that pathological vasoconstriction can increase cerebrovascular resistance and alter contrast transport despite proximal recanalization~\cite{Sun2024No-reflowStrategies., Kanoke2020TheStroke, Freitas-Andrade2020StructuralStroke}, leading to elevated arterial-phase signal and steeper slopes on DSA. Accordingly, early temporal features may reflect upstream hemodynamic effects associated with microvascular dysfunction and no-reflow~\cite{Qiu2023ARecanalization,Cipolla2014PostischemicArterioles}.

Within \texttt{TSF} (\texttt{SIPS + FLOW}), features describing post-peak dynamics and intensity distribution - most notably \texttt{plateauDuration}, \texttt{skewness}, \texttt{meanIntensity}, and \texttt{timeTo50Max} - were consistently ranked highest. These features reflect delayed washout, reduced contrast dispersion, and prolonged plateau phases, all hallmarks of impaired microvascular reperfusion. Lower importance was assigned to higher-order moments and variance-based descriptors, indicating limited incremental value beyond dominant temporal and intensity features. Importantly, when all feature categories were combined in the final model (\autoref{tab:model_performance}), its focus shifted almost entirely to \texttt{RAW}, with minimal reliance on \texttt{TSF} or \texttt{CLN} inputs, reflecting the complementary but derivative nature of \texttt{TSF}. These findings support the physiological interpretability of the model and reinforce the central role of post-EVT contrast dynamics in characterizing no-reflow.

\subsection{Feature ablations}
We systematically evaluated combinations of the \texttt{PEAK}, \texttt{SIPS}, and \texttt{FLOW} feature groups for no-reflow prediction (\autoref{tab:ablations}). Among individual feature sets, \texttt{FLOW} features demonstrated the strongest performance when using combined pre- and post-procedure signals (AUC = 0.8616), suggesting that temporal perfusion dynamics capture salient information related to microvascular reperfusion. \texttt{SIPS} features also performed well across configurations, particularly when derived from pre-procedure signals alone (AUC = 0.8080), highlighting the utility of global intensity-based descriptors. In contrast, \texttt{PEAK} features showed more modest and variable performance across settings. When evaluating pairwise combinations, the \texttt{SIPS + FLOW} feature set consistently achieved strong performance, yielding the highest AUC when using post-procedure features alone (AUC = 0.9018) and robust performance with pre-procedure features (AUC = 0.8348). Incorporating all three feature groups did not further improve performance, suggesting potential redundancy. Thus, the post-only \texttt{SIPS + FLOW} set (\texttt{TSF}) was selected for further analysis. Overall, models leveraging post-procedure features generally outperformed pre-procedure features alone, underscoring the importance of EVT-induced perfusion changes for no-reflow prediction.

\subsection{Subgroup analyses}
\label{subsection:sub_analyses}

We performed subgroup analyses of the best-performing combined model (\texttt{CLN+TSF+RAW}) to evaluate potential demographic and clinical confounding effects (\autoref{tab:sub_analysis}). Stratification by angiographic reperfusion status demonstrated comparable discrimination between patients with mTICI 2c (AUROC 0.96, $n=19$) and mTICI 3 (AUROC 0.89, $n=20$). Sex-based analyses showed similarly strong performance in females (AUROC 0.93, $n=28$) and males (AUROC 0.94, $n=11$), while race-based stratification yielded consistent AUROCs for White (0.92, $n=20$) and non-White patients (0.95, $n=19$). Age-based subgrouping around the cohort median of 78 years revealed similar performance in older patients (AUROC 0.97, $n=19$) compared with younger patients (AUROC 0.89, $n=20$). Stratification by baseline stroke severity (NIHSS $>$15 vs. $\leq$15) and number of thrombectomy passes (1 vs. $>$1) likewise demonstrated comparable AUROC values across groups. Subgroup differences in model behavior were assessed using two-sided Mann-Whitney U tests on continuous model output scores, stratified by true outcome to control for class prevalence. None of the observed subgroup differences reached statistical significance (all $p > 0.1$), indicating that model performance was consistent across demographic and clinical subgroups within this cohort, with no evidence of systematic bias.

\section{Discussion}
Our pilot preliminary results successfully demonstrate that pre- and post-procedure DSA sequences acquired during EVT provide sufficient information to predict no-reflow status towards the end of the procedure. The combination of all features (\texttt{CLN+TSF+RAW}), ensembling both lateral and AP views, enabled the best model to achieve an AUROC of $0.9330$, significantly outperforming the clinical baseline (AUROC: $0.7768$, $p=0.006$). No-reflow was defined as a $>15\%$ reduction in relative cerebral blood volume or flow within the infarct core compared to the contralateral hemisphere~\cite{ng_prevalence_2022}, using pre-EVT and up to 24-hour post-EVT MRI, supporting the feasibility of real-time angiographic prediction. Early identification of no-reflow risk could enable prompt risk stratification and post-EVT management, guiding interventions such as neurocritical care~\cite{sheriff_dynamic_2020, migdady_current_2023}, blood pressure optimization~\cite{peng_blood_2021, dong_blood_2024}, adjunctive aspiration~\cite{svilaas_thrombus_2008}, or intra-arterial thrombolysis~\cite{kaesmacher_safety_2020, renu_effect_2022}. Real-time model deployment could provide operators with automated risk estimates while still in the angiography suite, enabling immediate adjustments in procedural strategy or the initiation of adjunctive therapies.

The most predictive DSA-derived features capture fine-grained contrast intensity dynamics and distributional properties within downstream tissues, reflecting physiological signatures of microvascular impairment in no-reflow. In particular, temporal descriptors of contrast inflow and washout (\texttt{FLOW}), together with statistical intensity features (\texttt{SIPS}), characterize delayed clearance, reduced dispersion, and contrast stasis, that are established markers of impaired microvascular reperfusion and tissue-level hypoperfusion~\cite{mujanovic_no-reflow_2024, kan_no-reflow_2023, horie_recanalization_2025, schiphorst_tissue_2021}. With minimal human input (currently ~5-10 minutes, much lower with prospective automation) and fully automated downstream processing, integrating our model into clinical workflow would enable real-time identification of no-reflow immediately after EVT, a capability lacking in current practice. \emph{DSA-NRP} detects 100\% of no-reflow cases that would otherwise go unrecognized, allowing for rapid initiation of targeted interventions such as intensified neurocritical care or adjunctive intra-arterial thrombolysis. With the false-negative rate being 0\%, our approach substantially improves current clinical practice by identifying patients who would otherwise receive no targeted intervention under existing standards.

This pilot study has several limitations. The primary limitation is the relatively small cohort size of 39 patients from a single academic medical center, reflecting the strict inclusion of mTICI 2c/3 cases to ensure methodological rigor and cohort homogeneity. While this approach strengthens internal validity, it limits statistical power and generalizability, and increases the risk of overfitting despite the use of rigorous cross-validation. Accordingly, these findings should be interpreted as preliminary and require validation in larger, multi-center cohorts. Furthermore, although our feature importance analysis provides physiologically plausible interpretations, additional experimental and imaging studies are needed to more definitively establish the biological significance of the identified predictors. The study focused exclusively on M1 and ICA occlusions; extending to distal segments such as M2 branches would require modified TDT annotations. Inclusion of both pre- and post-procedure MRI for label generation may introduce selection bias, though MRI is routinely performed at our institution. Future work incorporating CT-based imaging and procedural metadata may improve generalizability and model robustness.

\section{Conclusion}
This study presents a novel framework for predicting no-reflow immediately after EVT using DSA sequences. By extracting structured features from both pre- and post-procedure DSA videos, we show that microvascular perfusion information can be captured at the time of the procedure, eliminating the need to wait for follow-up MRI. Models built on DSA-derived features substantially outperform those using clinical variables alone, underscoring the limited utility of routine clinical data. The use of interpretable, low-complexity ML models highlights this approach’s practicality for real-time implementation. Our preliminary findings establish the feasibility of predicting no-reflow from intra-procedural DSA alone, enabling early identification of high-risk patients and timely risk stratification during or after EVT. This could allow clinicians to adjust treatment decisions, such as individualized blood pressure management, adjunctive thrombus aspiration, or intra-arterial thrombolysis. Additionally, the model’s ability to generate risk estimates in real time during angiography highlights its potential for seamless deployment within existing procedural workflows. This work is an important step toward improving risk stratification and interventions for patients at risk of poor tissue reperfusion after EVT, ultimately aiming to improve long-term functional recovery.

\section*{Funding, Author Contributions and Acknowledgments}
This work was supported by the \textbf{National Institute of Neurological Disorders and Stroke, NIH (award R01NS100806)}. The authors have no conflicts of interest to disclose. 

Shreeram Athreya contributed to dataset curation, system design, and data analysis and interpretation. Carlos Olivares contributed to dataset curation and theoretical development. Both contributed equally to drafting, reviewing, and revising the article. Ameera Ismail provided clinical expertise, contributed to dataset curation, annotation, and theoretical development, and participated in manuscript preparation. Kambiz Nael provided clinical expertise, contributed to theoretical development, and manuscript revision. William Speier contributed to data analysis, interpretation, and project oversight, and participated in drafting and revising the article. Corey Arnold provided project oversight, secured funding, and contributed to manuscript preparation. All authors reviewed and approved the final manuscript.

The authors would like to thank Santi Bhattarai-Kline (DGSOM UCLA) for helpful discussions on physiological relevance. We also thank Shawn Chen (UCLA Radiology) for helping us with data collection.


\bibliographystyle{ama.bst}
\bibliography{main.bib}

@article{de2024robust,
  title={A robust ensemble algorithm for ischemic stroke lesion segmentation: Generalizability and clinical utility beyond the isles challenge},
  author={de la Rosa, Ezequiel and Reyes, Mauricio and Liew, Sook-Lei and Hutton, Alexandre and Wiest, Roland and Kaesmacher, Johannes and Hanning, Uta and Hakim, Arsany and Zubal, Richard and Valenzuela, Waldo and others},
  journal={arXiv preprint arXiv:2403.19425},
  year={2024}
}

@article{powers_2015_2015,
    title = {2015 {American} {Heart} {Association}/{American} {Stroke} {Association} {Focused} {Update} of the 2013 {Guidelines} for the {Early} {Management} of {Patients} {With} {Acute} {Ischemic} {Stroke} {Regarding} {Endovascular} {Treatment}: {A} {Guideline} for {Healthcare} {Professionals} {From} the {American} {Heart} {Association}/{American} {Stroke} {Association}},
    volume = {46},
    issn = {0039-2499, 1524-4628},
    shorttitle = {2015 {American} {Heart} {Association}/{American} {Stroke} {Association} {Focused} {Update} of the 2013 {Guidelines} for the {Early} {Management} of {Patients} {With} {Acute} {Ischemic} {Stroke} {Regarding} {Endovascular} {Treatment}},
    doi = {10.1161/STR.0000000000000074},
    language = {en},
    number = {10},
    urldate = {2025-03-17},
    journal = {Stroke},
    author = {Powers, William J. and Derdeyn, Colin P. and Biller, José and Coffey, Christopher S. and Hoh, Brian L. and Jauch, Edward C. and Johnston, Karen C. and Johnston, S. Claiborne and Khalessi, Alexander A. and Kidwell, Chelsea S. and Meschia, James F. and Ovbiagele, Bruce and Yavagal, Dileep R.},
    month = oct,
    year = {2015},
    pages = {3020--3035},
}

@article{cho_mechanical_2021,
    title = {Mechanical thrombectomy for acute ischemic stroke with occlusion of the {M2} segment of the middle cerebral artery: {A} literature review},
    volume = {23},
    issn = {2234-8565, 2287-3139},
    shorttitle = {Mechanical thrombectomy for acute ischemic stroke with occlusion of the {M2} segment of the middle cerebral artery},
    doi = {10.7461/jcen.2021.E2020.11.002},
    language = {en},
    number = {3},
    urldate = {2025-03-17},
    journal = {Journal of Cerebrovascular and Endovascular Neurosurgery},
    author = {Cho, Yong-Hwan and Choi, Jae Hyung},
    month = sep,
    year = {2021},
    pages = {193--200},
}

@article{liyis_mechanical_2023,
    title = {Mechanical thrombectomy in {M1} and {M2} segments of middle cerebral arteries: {A} meta-analysis of prospective cohort studies},
    volume = {231},
    issn = {03038467},
    shorttitle = {Mechanical thrombectomy in {M1} and {M2} segments of middle cerebral arteries},
    doi = {10.1016/j.clineuro.2023.107823},
    language = {en},
    urldate = {2025-03-17},
    journal = {Clinical Neurology and Neurosurgery},
    author = {Liyis, Bryan Gervais De and Surya, Stevanus Christian and Tedyanto, Eric Hartono and Pramana, Nyoman Angga Krishna and Widyadharma, I. Putu Eka},
    month = aug,
    year = {2023},
    pages = {107823},
}

@article{mujanovic_no-reflow_2024,
    title = {No-reflow phenomenon in stroke patients: {A} systematic literature review and meta-analysis of clinical data},
    volume = {19},
    issn = {1747-4930, 1747-4949},
    shorttitle = {No-reflow phenomenon in stroke patients},
    doi = {10.1177/17474930231180434},
    language = {en},
    number = {1},
    urldate = {2025-03-17},
    journal = {International Journal of Stroke},
    author = {Mujanovic, Adnan and Ng, Felix and Meinel, Thomas R and Dobrocky, Tomas and Piechowiak, Eike I and Kurmann, Christoph C and Seiffge, David J and Wegener, Susanne and Wiest, Roland and Meyer, Lukas and Fiehler, Jens and Olivot, Jean Marc and Ribo, Marc and Nguyen, Thanh N and Gralla, Jan and Campbell, Bruce Cv and Fischer, Urs and Kaesmacher, Johannes},
    month = jan,
    year = {2024},
    pages = {58--67},
}

@article{horie_recanalization_2025,
    title = {Recanalization {Does} {Not} {Always} {Equate} to {Reperfusion}: {No}-{Reflow} {Phenomenon} {After} {Successful} {Thrombectomy}},
    volume = {56},
    issn = {0039-2499, 1524-4628},
    shorttitle = {Recanalization {Does} {Not} {Always} {Equate} to {Reperfusion}},
    doi = {10.1161/STROKEAHA.124.048994},
    language = {en},
    number = {1},
    urldate = {2025-03-17},
    journal = {Stroke},
    author = {Horie, Nobutaka and Inoue, Manabu and Morimoto, Takeshi and Sadakata, Eisaku and Okamura, Kazuaki and Morofuji, Yoichi and Hara, Takeshi and Kuwabara, Masashi and Kondo, Hiroshi and Ishii, Daizo},
    month = jan,
    year = {2025},
    pages = {183--189},
}

@article{ng_prevalence_2022,
    title = {Prevalence and {Significance} of {Impaired} {Microvascular} {Tissue} {Reperfusion} {Despite} {Macrovascular} {Angiographic} {Reperfusion} ({No}-{Reflow})},
    volume = {98},
    issn = {0028-3878, 1526-632X},
    doi = {10.1212/WNL.0000000000013210},
    language = {en},
    number = {8},
    urldate = {2025-03-17},
    journal = {Neurology},
    author = {Ng, Felix C. and Churilov, Leonid and Yassi, Nawaf and Kleinig, Timothy John and Thijs, Vincent and Wu, Teddy and Shah, Darshan and Dewey, Helen and Sharma, Gagan and Desmond, Patricia and Yan, Bernard and Parsons, Mark and Donnan, Geoffrey and Davis, Stephen and Mitchell, Peter and Campbell, Bruce},
    month = feb,
    year = {2022},
}

@article{mujanovic_association_2023,
    title = {Association of {Intravenous} {Thrombolysis} with {Delayed} {Reperfusion} {After} {Incomplete} {Mechanical} {Thrombectomy}},
    volume = {33},
    issn = {1869-1439, 1869-1447},
    doi = {10.1007/s00062-022-01186-7},
    language = {en},
    number = {1},
    urldate = {2025-03-17},
    journal = {Clinical Neuroradiology},
    author = {Mujanovic, Adnan and Kammer, Christoph and Kurmann, Christoph C. and Grunder, Lorenz and Beyeler, Morin and Lang, Matthias F. and Piechowiak, Eike I. and Meinel, Thomas R. and Jung, Simon and Almiri, William and Pilgram-Pastor, Sara and Hoffmann, Angelika and Seiffge, David J. and Heldner, Mirjam R. and Dobrocky, Tomas and Mordasini, Pasquale and Arnold, Marcel and Gralla, Jan and Fischer, Urs and Kaesmacher, Johannes},
    month = mar,
    year = {2023},
    pages = {87--98},
}

@article{liebeskind_etici_2019,
    title = {{eTICI} reperfusion: defining success in endovascular stroke therapy},
    volume = {11},
    issn = {1759-8478, 1759-8486},
    shorttitle = {{eTICI} reperfusion},
    doi = {10.1136/neurintsurg-2018-014127},
    language = {en},
    number = {5},
    urldate = {2025-03-17},
    journal = {Journal of NeuroInterventional Surgery},
    author = {Liebeskind, David S and Bracard, Serge and Guillemin, Francis and Jahan, Reza and Jovin, Tudor G and Majoie, Charles Blm and Mitchell, Peter J and Van Der Lugt, Aad and Menon, Bijoy K and San Román, Luis and Campbell, Bruce Cv and Muir, Keith W and Hill, Michael D and Dippel, Diederik Wj and Saver, Jeffrey L and Demchuk, Andrew M and Dávalos, Antoni and White, Philip and Brown, Scott and Goyal, Mayank},
    month = may,
    year = {2019},
    pages = {433--438},
}

@article{abdalkader_neuroimaging_2023,
    title = {Neuroimaging of {Acute} {Ischemic} {Stroke}: {Multimodal} {Imaging} {Approach} for {Acute} {Endovascular} {Therapy}},
    volume = {25},
    issn = {2287-6391, 2287-6405},
    shorttitle = {Neuroimaging of {Acute} {Ischemic} {Stroke}},
    doi = {10.5853/jos.2022.03286},
    language = {en},
    number = {1},
    urldate = {2025-03-17},
    journal = {Journal of Stroke},
    author = {Abdalkader, Mohamad and Siegler, James E. and Lee, Jin Soo and Yaghi, Shadi and Qiu, Zhongming and Huo, Xiaochuan and Miao, Zhongrong and Campbell, Bruce C.V. and Nguyen, Thanh N.},
    month = jan,
    year = {2023},
    pages = {55--71},
}

@article{hwang_comparative_2012,
    title = {Comparative {Sensitivity} of {Computed} {Tomography} vs. {Magnetic} {Resonance} {Imaging} for {Detecting} {Acute} {Posterior} {Fossa} {Infarct}},
    volume = {42},
    copyright = {https://www.elsevier.com/tdm/userlicense/1.0/},
    issn = {07364679},
    doi = {10.1016/j.jemermed.2011.05.101},
    language = {en},
    number = {5},
    urldate = {2025-03-17},
    journal = {The Journal of Emergency Medicine},
    author = {Hwang, David Y. and Silva, Gisele S. and Furie, Karen L. and Greer, David M.},
    month = may,
    year = {2012},
    pages = {559--565},
}

@article{kan_no-reflow_2023,
    title = {No-reflow phenomenon following stroke recanalization therapy: {Clinical} assessment advances: {A} narrative review},
    volume = {9},
    issn = {2394-8108, 2455-4626},
    shorttitle = {No-reflow phenomenon following stroke recanalization therapy},
    doi = {10.4103/bc.bc_37_23},
    language = {en},
    number = {4},
    urldate = {2025-03-17},
    journal = {Brain Circulation},
    author = {Kan, Yuan and Li, Sijie and Zhang, Bowei and Ding, Yuchuan and Zhao, Wenbo and Ji, Xunming},
    month = oct,
    year = {2023},
    pages = {214--221},
}

@article{arnould_comparison_2004,
    title = {Comparison of {CT} and three {MR} sequences for detecting and categorizing early (48 hours) hemorrhagic transformation in hyperacute ischemic stroke},
    volume = {25},
    issn = {0195-6108},
    language = {eng},
    number = {6},
    journal = {AJNR. American journal of neuroradiology},
    author = {Arnould, Marie-Cécile and Grandin, Cécile B. and Peeters, André and Cosnard, Guy and Duprez, Thierry P.},
    year = {2004},
    pmid = {15205127},
    pmcid = {PMC7975678},
    pages = {939--944},
}

@article{huang_comprehensive_2024,
    title = {A {Comprehensive} {Prediction} {Model} for {Futile} {Recanalization} in {AIS} {Patients} {Post}-{Endovascular} {Therapy}: {Integrating} {Clinical}, {Imaging}, and {No}-{Reflow}  {Biomarkers}.},
    volume = {15},
    issn = {2152-5250},
    doi = {10.14336/AD.2024.0127},
    language = {eng},
    number = {6},
    journal = {Aging and disease},
    author = {Huang, Shuangfeng and Xu, Jiali and Kang, Haijuan and Guo, Wenting and Ren, Changhong and Wehbe, Alexandra and Song, Haiqing and Ma, Qingfeng and Zhao, Wenbo and Ding, Yuchuan and Ji, Xunming and Li, Sijie},
    month = apr,
    year = {2024},
    pmid = {38739941},
    pmcid = {PMC11567269},
    note = {Place: United States},
    pages = {2852--2862},
}

@article{nicolini_no-reflow_2023,
    title = {No-reflow phenomenon in acute ischemic stroke: an angiographic evaluation},
    volume = {44},
    issn = {1590-1874, 1590-3478},
    shorttitle = {No-reflow phenomenon in acute ischemic stroke},
    doi = {10.1007/s10072-023-06879-6},
    language = {en},
    number = {11},
    urldate = {2025-03-18},
    journal = {Neurological Sciences},
    author = {Nicolini, Ettore and Iacobucci, Marta and De Michele, Manuela and Ciacciarelli, Antonio and Berto, Irene and Petraglia, Luca and Falcou, Anne and Cirelli, Carlo and Biraschi, Francesco and Lorenzano, Svetlana and Linfante, Italo and Toni, Danilo},
    month = nov,
    year = {2023},
    pages = {3939--3948},
}

@article{prasetya_qtici_2021,
    title = {{qTICI}: {Quantitative} assessment of brain tissue reperfusion on digital subtraction angiograms of acute ischemic stroke patients},
    volume = {16},
    issn = {1747-4930, 1747-4949},
    shorttitle = {{qTICI}},
    doi = {10.1177/1747493020909632},
    language = {en},
    number = {2},
    urldate = {2025-03-18},
    journal = {International Journal of Stroke},
    author = {Prasetya, Haryadi and Ramos, Lucas A and Epema, Thabiso and Treurniet, Kilian M and Emmer, Bart J and Van Den Wijngaard, Ido R and Zhang, Guang and Kappelhof, Manon and Berkhemer, Olvert A and Yoo, Albert J and Roos, Yvo Bewm and Van Oostenbrugge, Robert J and Dippel, Diederik Wj and Van Zwam, Wim H and Van Der Lugt, Aad and De Mol, Bas Ajm and Majoie, Charles Blm and Bavel, Ed Van and Marquering, Henk A and {on behalf of the MR CLEAN Registry Investigators}},
    month = feb,
    year = {2021},
    pages = {207--216},
}

@article{sun_predictors_2023,
    title = {Predictors of futile recanalization after endovascular treatment in acute ischemic stroke: a multi-center study},
    volume = {17},
    issn = {1662-453X},
    shorttitle = {Predictors of futile recanalization after endovascular treatment in acute ischemic stroke},
    doi = {10.3389/fnins.2023.1279366},
    urldate = {2025-03-18},
    journal = {Frontiers in Neuroscience},
    author = {Sun, Yu and Jou, Eric and Nguyen, Thanh N. and Mofatteh, Mohammad and Liang, Qingjia and Abdalkader, Mohamad and Yan, Zile and Feng, Mingzhu and Li, Xinyuan and Li, Guilan and Luo, Lanzhu and Lai, Yuzheng and Yang, Shuiquan and Zhou, Sijie and Xu, Zhiming and Cai, Xiaodong and Chen, Yimin},
    month = nov,
    year = {2023},
    pages = {1279366},
}

@article{sabieleish_image_2021,
    title = {Image processing-based {mTICI} grading after endovascular treatment for acute ischemic stroke},
    volume = {7},
    copyright = {http://creativecommons.org/licenses/by/4.0},
    issn = {2364-5504},
    doi = {10.1515/cdbme-2021-2060},
    language = {en},
    number = {2},
    urldate = {2025-03-20},
    journal = {Current Directions in Biomedical Engineering},
    author = {Sabieleish, Muhannad and Thormann, Maximilian and Metzler, Jonathan and Boese, Axel and Friebe, Michael and Mpotsaris, Anastasios and Behme, Daniel},
    month = oct,
    year = {2021},
    pages = {235--238},
}

@article{mittmann_deep_2022,
    title = {Deep learning-based classification of {DSA} image sequences of patients with acute ischemic stroke},
    volume = {17},
    issn = {1861-6429},
    doi = {10.1007/s11548-022-02654-8},
    language = {en},
    number = {9},
    urldate = {2025-03-20},
    journal = {International Journal of Computer Assisted Radiology and Surgery},
    author = {Mittmann, Benjamin J. and Braun, Michael and Runck, Frank and Schmitz, Bernd and Tran, Thuy N. and Yamlahi, Amine and Maier-Hein, Lena and Franz, Alfred M.},
    month = may,
    year = {2022},
    pages = {1633--1641},
}

@inproceedings{zhang_machine_2021,
    address = {Athens, Greece},
    title = {A {Machine} {Learning} {Approach} to {Predict} {Acute} {Ischemic} {Stroke} {Thrombectomy} {Reperfusion} using {Discriminative} {MR} {Image} {Features}},
    copyright = {https://doi.org/10.15223/policy-029},
    isbn = {978-1-6654-0358-0},
    doi = {10.1109/BHI50953.2021.9508597},
    urldate = {2025-03-20},
    booktitle = {2021 {IEEE} {EMBS} {International} {Conference} on {Biomedical} and {Health} {Informatics} ({BHI})},
    publisher = {IEEE},
    author = {Zhang, Haoyue and Polson, Jennifer and Nael, Kambiz and Salamon, Noriko and Yoo, Bryan and Speier, William and Arnold, Corey},
    month = jul,
    year = {2021},
    pages = {1--4},
}

@article{da_ros_ensemble_2024,
    title = {Ensemble machine learning to predict futile recanalization after mechanical thrombectomy based on non-contrast {CT} imaging},
    volume = {33},
    issn = {10523057},
    doi = {10.1016/j.jstrokecerebrovasdis.2024.107890},
    language = {en},
    number = {11},
    urldate = {2025-03-20},
    journal = {Journal of Stroke and Cerebrovascular Diseases},
    author = {Da Ros, Valerio and Cavallo, Armando and Di Donna, Carlo and D'Onofrio, Adolfo and Trulli, Mariafrancesca and Di Candia, Simone and Mancini, Ludovica and Funari, Luca and Cecchi, Gianluca and Carini, Alessandro and Madonna, Matteo and Sabuzi, Federico and Di Giuliano, Francesca and Zelenak, Kamil and Diomedi, Marina and Maestrini, Ilaria and Garaci, Francesco},
    month = nov,
    year = {2024},
    pages = {107890},
}

@article{kelly_deep_2023,
    title = {{DEEP} {MOVEMENT}: {Deep} learning of movie files for management of endovascular thrombectomy},
    volume = {33},
    issn = {1432-1084},
    shorttitle = {{DEEP} {MOVEMENT}},
    doi = {10.1007/s00330-023-09478-3},
    language = {en},
    number = {8},
    urldate = {2025-03-20},
    journal = {European Radiology},
    author = {Kelly, Brendan and Martinez, Mesha and Do, Huy and Hayden, Joel and Huang, Yuhao and Yedavalli, Vivek and Ho, Chang and Keane, Pearse A. and Killeen, Ronan and Lawlor, Aonghus and Moseley, Michael E. and Yeom, Kristen W. and Lee, Edward H.},
    month = feb,
    year = {2023},
    pages = {5728--5739},
}

@article{xu_cvfsnet_2025,
    title = {{CVFSNet}: {A} {Cross} {View} {Fusion} {Scoring} {Network} for end-to-end {mTICI} scoring},
    volume = {102},
    issn = {13618415},
    shorttitle = {{CVFSNet}},
    doi = {10.1016/j.media.2025.103508},
    language = {en},
    urldate = {2025-03-20},
    journal = {Medical Image Analysis},
    author = {Xu, Weijin and Tan, Tao and Yang, Huihua and Liu, Wentao and Chen, Yifu and Zhang, Ling and Pan, Xipeng and Gao, Feng and Deng, Yiming and Van Walsum, Theo and Van Der Sluijs, Matthijs and Su, Ruisheng},
    month = may,
    year = {2025},
    pages = {103508},
}

@article{saini_global_2021,
    title = {Global {Epidemiology} of {Stroke} and {Access} to {Acute} {Ischemic} {Stroke} {Interventions}},
    volume = {97},
    issn = {0028-3878, 1526-632X},
    doi = {10.1212/WNL.0000000000012781},
    language = {en},
    number = {20\_Supplement\_2},
    urldate = {2025-03-25},
    journal = {Neurology},
    author = {Saini, Vasu and Guada, Luis and Yavagal, Dileep R.},
    month = nov,
    year = {2021},
}

@article{kim_global_2024,
    title = {Global stroke statistics 2023: {Availability} of reperfusion services around the world},
    volume = {19},
    issn = {1747-4930, 1747-4949},
    shorttitle = {Global stroke statistics 2023},
    doi = {10.1177/17474930231210448},
    language = {en},
    number = {3},
    urldate = {2025-03-25},
    journal = {International Journal of Stroke},
    author = {Kim, Joosup and Olaiya, Muideen T and De Silva, Deidre A and Norrving, Bo and Bosch, Jackie and De Sousa, Diana A and Christensen, Hanne K and Ranta, Anna and Donnan, Geoffrey A and Feigin, Valery and Martins, Sheila and Schwamm, Lee H and Werring, David J and Howard, George and Owolabi, Mayowa and Pandian, Jeyaraj and Mikulik, Robert and Thayabaranathan, Tharshanah and Cadilhac, Dominique A},
    month = mar,
    year = {2024},
    pages = {253--270},
}

@article{karonen_combined_1999,
    title = {Combined {Diffusion} and {Perfusion} {MRI} {With} {Correlation} to {Single}-{Photon} {Emission} {CT} in {Acute} {Ischemic} {Stroke}: {Ischemic} {Penumbra} {Predicts} {Infarct} {Growth}},
    volume = {30},
    issn = {0039-2499, 1524-4628},
    shorttitle = {Combined {Diffusion} and {Perfusion} {MRI} {With} {Correlation} to {Single}-{Photon} {Emission} {CT} in {Acute} {Ischemic} {Stroke}},
    doi = {10.1161/01.STR.30.8.1583},
    language = {en},
    number = {8},
    urldate = {2025-03-25},
    journal = {Stroke},
    author = {Karonen, Jari O. and Vanninen, Ritva L. and Liu, Yawu and Østergaard, Leif and Kuikka, Jyrki T. and Nuutinen, Juho and Vanninen, Esko J. and Partanen, P. L. Kaarina and Vainio, Pauli A. and Korhonen, Katja and Perkiö, Jussi and Roivainen, Reina and Sivenius, Juhani and Aronen, Hannu J.},
    month = aug,
    year = {1999},
    pages = {1583--1590},
}

@article{badhiwala_endovascular_2015,
    title = {Endovascular {Thrombectomy} for {Acute} {Ischemic} {Stroke}: {A} {Meta}-analysis},
    volume = {314},
    issn = {0098-7484},
    shorttitle = {Endovascular {Thrombectomy} for {Acute} {Ischemic} {Stroke}},
    doi = {10.1001/jama.2015.13767},
    language = {en},
    number = {17},
    urldate = {2025-03-25},
    journal = {JAMA},
    author = {Badhiwala, Jetan H. and Nassiri, Farshad and Alhazzani, Waleed and Selim, Magdy H. and Farrokhyar, Forough and Spears, Julian and Kulkarni, Abhaya V. and Singh, Sheila and Alqahtani, Abdulrahman and Rochwerg, Bram and Alshahrani, Mohammad and Murty, Naresh K. and Alhazzani, Adel and Yarascavitch, Blake and Reddy, Kesava and Zaidat, Osama O. and Almenawer, Saleh A.},
    month = nov,
    year = {2015},
    pages = {1832},
}

@article{ganesh_thrombectomy_2018,
    title = {Thrombectomy for {Acute} {Ischemic} {Stroke}: {Recent} {Insights} and {Future} {Directions}},
    volume = {18},
    issn = {1528-4042, 1534-6293},
    shorttitle = {Thrombectomy for {Acute} {Ischemic} {Stroke}},
    doi = {10.1007/s11910-018-0869-8},
    language = {en},
    number = {9},
    urldate = {2025-03-25},
    journal = {Current Neurology and Neuroscience Reports},
    author = {Ganesh, Aravind and Goyal, Mayank},
    month = sep,
    year = {2018},
    pages = {59},
}

@article{schiphorst_tissue_2021,
    title = {Tissue \textit{no-reflow} despite full recanalization following thrombectomy for anterior circulation stroke with proximal occlusion: {A} clinical study},
    volume = {41},
    issn = {0271-678X, 1559-7016},
    shorttitle = {Tissue \textit{no-reflow} despite full recanalization following thrombectomy for anterior circulation stroke with proximal occlusion},
    doi = {10.1177/0271678X20954929},
    language = {en},
    number = {2},
    urldate = {2025-03-27},
    journal = {Journal of Cerebral Blood Flow \& Metabolism},
    author = {Schiphorst, Adrien Ter and Charron, Sylvain and Hassen, Wagih Ben and Provost, Corentin and Naggara, Olivier and Benzakoun, Joseph and Seners, Pierre and Turc, Guillaume and Baron, Jean-Claude and Oppenheim, Catherine},
    month = feb,
    year = {2021},
    pages = {253--266},
}

@article{patel_hyperacute_2020,
    title = {Hyperacute {Management} of {Ischemic} {Strokes}},
    volume = {75},
    issn = {07351097},
    doi = {10.1016/j.jacc.2020.03.006},
    language = {en},
    number = {15},
    urldate = {2025-04-02},
    journal = {Journal of the American College of Cardiology},
    author = {Patel, Pratit and Yavagal, Dileep and Khandelwal, Priyank},
    month = apr,
    year = {2020},
    pages = {1844--1856},
}

@article{dong_blood_2024,
    title = {Blood pressure management after endovascular thrombectomy: {Insights} of recent randomized controlled trials},
    volume = {30},
    issn = {1755-5930, 1755-5949},
    shorttitle = {Blood pressure management after endovascular thrombectomy},
    doi = {10.1111/cns.14907},
    language = {en},
    number = {8},
    urldate = {2025-04-05},
    journal = {CNS Neuroscience \& Therapeutics},
    author = {Dong, Xiao and Liu, Yuanyuan and Chu, Xuehong and Yu, Erlan and Jia, Xiaole and Ji, Xunming and Wu, Chuanjie},
    month = aug,
    year = {2024},
    pages = {e14907},
}

@article{peng_blood_2021,
    title = {Blood {Pressure} {Management} {After} {Endovascular} {Thrombectomy}},
    volume = {12},
    issn = {1664-2295},
    doi = {10.3389/fneur.2021.723461},
    urldate = {2025-04-05},
    journal = {Frontiers in Neurology},
    author = {Peng, Teng J. and Ortega-Gutiérrez, Santiago and De Havenon, Adam and Petersen, Nils H.},
    month = sep,
    year = {2021},
    pages = {723461},
}

@article{sheriff_dynamic_2020,
    title = {Dynamic {Cerebral} {Autoregulation} {Post} {Endovascular} {Thrombectomy} in {Acute} {Ischemic} {Stroke}},
    volume = {10},
    copyright = {https://creativecommons.org/licenses/by/4.0/},
    issn = {2076-3425},
    doi = {10.3390/brainsci10090641},
    language = {en},
    number = {9},
    urldate = {2025-04-07},
    journal = {Brain Sciences},
    author = {Sheriff, Faheem and Castro, Pedro and Kozberg, Mariel and LaRose, Sarah and Monk, Andrew and Azevedo, Elsa and Li, Karen and Jafari, Sameen and Rao, Shyam and Otite, Fadar Oliver and Khawaja, Ayaz and Sorond, Farzaneh and Feske, Steven and Tan, Can Ozan and Vaitkevicius, Henrikas},
    month = sep,
    year = {2020},
    pages = {641},
}

@article{migdady_current_2023,
    title = {Current and {Emerging} {Endovascular} and {Neurocritical} {Care} {Management} {Strategies} in {Large}-{Core} {Ischemic} {Stroke}},
    volume = {12},
    copyright = {https://creativecommons.org/licenses/by/4.0/},
    issn = {2077-0383},
    doi = {10.3390/jcm12206641},
    language = {en},
    number = {20},
    urldate = {2025-04-07},
    journal = {Journal of Clinical Medicine},
    author = {Migdady, Ibrahim and Johnson-Black, Phoebe H. and Leslie-Mazwi, Thabele and Malhotra, Rishi},
    month = oct,
    year = {2023},
    pages = {6641},
}

@article{svilaas_thrombus_2008,
    title = {Thrombus {Aspiration} during {Primary} {Percutaneous} {Coronary} {Intervention}},
    volume = {358},
    issn = {0028-4793, 1533-4406},
    doi = {10.1056/NEJMoa0706416},
    language = {en},
    number = {6},
    urldate = {2025-04-07},
    journal = {New England Journal of Medicine},
    author = {Svilaas, Tone and Vlaar, Pieter J. and Van Der Horst, Iwan C. and Diercks, Gilles F.H. and De Smet, Bart J.G.L. and Van Den Heuvel, Ad F.M. and Anthonio, Rutger L. and Jessurun, Gillian A. and Tan, Eng-Shiong and Suurmeijer, Albert J.H. and Zijlstra, Felix},
    month = feb,
    year = {2008},
    pages = {557--567},
}

@article{kaesmacher_safety_2020,
    title = {Safety and {Efficacy} of {Intra}-arterial {Urokinase} {After} {Failed}, {Unsuccessful}, or {Incomplete} {Mechanical} {Thrombectomy} in {Anterior} {Circulation} {Large}-{Vessel} {Occlusion} {Stroke}},
    volume = {77},
    issn = {2168-6149},
    doi = {10.1001/jamaneurol.2019.4192},
    language = {en},
    number = {3},
    urldate = {2025-04-07},
    journal = {JAMA Neurology},
    author = {Kaesmacher, Johannes and Bellwald, Sebastian and Dobrocky, Tomas and Meinel, Thomas R. and Piechowiak, Eike I. and Goeldlin, Martina and Kurmann, Christoph C. and Heldner, Mirjam R. and Jung, Simon and Mordasini, Pasquale and Arnold, Marcel and Mosimann, Pascal J. and Schroth, Gerhard and Mattle, Heinrich P. and Gralla, Jan and Fischer, Urs},
    month = mar,
    year = {2020},
    pages = {318},
}

@article{renu_effect_2022,
    title = {Effect of {Intra}-arterial {Alteplase} vs {Placebo} {Following} {Successful} {Thrombectomy} on {Functional} {Outcomes} in {Patients} {With} {Large} {Vessel} {Occlusion} {Acute} {Ischemic} {Stroke}: {The} {CHOICE} {Randomized} {Clinical} {Trial}},
    volume = {327},
    issn = {0098-7484},
    shorttitle = {Effect of {Intra}-arterial {Alteplase} vs {Placebo} {Following} {Successful} {Thrombectomy} on {Functional} {Outcomes} in {Patients} {With} {Large} {Vessel} {Occlusion} {Acute} {Ischemic} {Stroke}},
    doi = {10.1001/jama.2022.1645},
    language = {en},
    number = {9},
    urldate = {2025-04-07},
    journal = {JAMA},
    author = {Renú, Arturo and Millán, Mónica and San Román, Luis and Blasco, Jordi and Martí-Fàbregas, Joan and Terceño, Mikel and Amaro, Sergio and Serena, Joaquín and Urra, Xabier and Laredo, Carlos and Barranco, Roger and Camps-Renom, Pol and Zarco, Federico and Oleaga, Laura and Cardona, Pere and Castaño, Carlos and Macho, Juan and Cuadrado-Godía, Elisa and Vivas, Elio and López-Rueda, Antonio and Guimaraens, Leopoldo and Ramos-Pachón, Anna and Roquer, Jaume and Muchada, Marian and Tomasello, Alejandro and Dávalos, Antonio and Torres, Ferran and Chamorro, Ángel and {CHOICE Investigators} and Llull, Laura and Vargas, Martha and Obach, Victor and Rudilosso, Salvatore and Rodríguez-Vázquez, Alejandro and Santana, Daniel and Macías, Napoleón and Serrano, Elena and Moreno, Javier and Pérez De La Ossa, Natalia and Dorado, Laura and Hernández-Pérez, Maria and Gomis, Meritxell and Muñoz, Lucia and Rodríguez-Molinos, P. and Boix, Martí and Palomeras, Ernest and Núñez, F. and Remollo, Sebastián and Werner, Mariano and Vera, Victor and Paul, Laura and Pardo, Laura and Reina, Montserrat and Bashir, Saima and Bojaryn, Ursula and Silva, Yolanda and Guasch, Marina and Murillo, Alan and Rodríguez Álvarez-Cienfuegos, Juan and Comas, M. and Martínez, B. and Nogué, E. and Chirife, Oscar and Quesada, Helena and Lara, Blanca and Paipa, Andres and Aja, Lucia and Mora, Paloma and De Miquel, Maria Angel and Aixut, Sónia and Ferrer, Anna Maria and Marín, R. and Prats-Sánchez, Luis and Delgado-Mederos, Raquel and Martínez-Domeño, A and Branera, Joan and Guerrero, R. and Villalba, , J and Rodríguez, , A and Berga, N. and Jiménez-Xarrié, Elena and Romeral, G. and Ois, Ángel and Jiménez, Jordi and Avellaneda, C. and Cayuela, , N and Rodríguez, Ana and Giralt, Eva and Espona, M and Saldaña, J. and Hernández, David and Ribó, Marc and Piñana, Carlos and Rodríguez, Noelia and Boned, Sandra and Molina, Carlos and Rubiera, Marta and Juega, Jesús and Rodríguez-Luna, David and Pagola, Juan and Garcia-Tonel, Alvaro and Deck, Matias and Sala, V and Sanjuan, Estela and Santana, K. and Losada, C. and Suñe, P. and Jovin, Tudor G. and Leira, Enrique and Rios, José},
    month = mar,
    year = {2022},
    pages = {826},
}

@article{zhang_deep_2024,
    title = {A {Deep} {Learning} {Approach} to {Predict} {Recanalization} {First}-{Pass} {Effect} following {Mechanical} {Thrombectomy} in {Patients} with {Acute} {Ischemic} {Stroke}},
    volume = {45},
    issn = {0195-6108, 1936-959X},
    doi = {10.3174/ajnr.A8272},
    language = {en},
    number = {8},
    urldate = {2025-05-22},
    journal = {American Journal of Neuroradiology},
    author = {Zhang, Haoyue and Polson, Jennifer S. and Wang, Zichen and Nael, Kambiz and Rao, Neal M. and Speier, William F. and Arnold, Corey W.},
    month = aug,
    year = {2024},
    pages = {1044--1052},
}

@article{su_autotici_2021,
    title = {{autoTICI}: {Automatic} {Brain} {Tissue} {Reperfusion} {Scoring} on {2D} {DSA} {Images} of {Acute} {Ischemic} {Stroke} {Patients}},
    volume = {40},
    copyright = {https://ieeexplore.ieee.org/Xplorehelp/downloads/license-information/IEEE.html},
    issn = {0278-0062, 1558-254X},
    shorttitle = {{autoTICI}},
    doi = {10.1109/TMI.2021.3077113},
    number = {9},
    urldate = {2025-06-06},
    journal = {IEEE Transactions on Medical Imaging},
    author = {Su, Ruisheng and Cornelissen, Sandra A. P. and Van Der Sluijs, Matthijs and Van Es, Adriaan C. G. M. and Van Zwam, Wim H. and Dippel, Diederik W. J. and Lycklama, Geert and Van Doormaal, Pieter Jan and Niessen, Wiro J. and Van Der Lugt, Aad and Van Walsum, Theo},
    month = sep,
    year = {2021},
    pages = {2380--2391},
}

@article{lee_automatic_2017,
    title = {Automatic flow analysis of digital subtraction angiography using independent component analysis in patients with carotid stenosis},
    volume = {12},
    issn = {1932-6203},
    doi = {10.1371/journal.pone.0185330},
    language = {en},
    number = {9},
    urldate = {2025-06-06},
    journal = {PLOS ONE},
    author = {Lee, Han-Jui and Hong, Jia-Sheng and Lin, Chung-Jung and Kao, Yi-Hsuan and Chang, Feng-Chi and Luo, Chao-Bao and Chu, Wei-Fa},
    editor = {Meckel, Stephan},
    month = sep,
    year = {2017},
    pages = {e0185330},
}

@article{lu_digital_2021,
    title = {Digital {Subtraction} {Angiography} {Contrast} {Material} {Transport} as a {Direct} {Assessment} for {Blood} {Perfusion} of {Middle} {Cerebral} {Artery} {Stenosis}},
    volume = {12},
    issn = {1664-042X},
    doi = {10.3389/fphys.2021.716173},
    urldate = {2025-06-06},
    journal = {Frontiers in Physiology},
    author = {Lu, Yun-Hao and Cai, Yan and Zhang, Yi and Wang, Rui and Li, Zhi-Yong},
    month = aug,
    year = {2021},
    pages = {716173},
}

@article{jann_implication_2016,
    title = {Implication of cerebral circulation time in intracranial stenosis measured by digital subtraction angiography on cerebral blood flow estimation measured by arterial spin labeling},
    volume = {22},
    issn = {13053825, 13053612},
    doi = {10.5152/dir.2016.15204},
    number = {5},
    urldate = {2025-06-06},
    journal = {Diagnostic and Interventional Radiology},
    author = {Jann, Kay and Hauf, Martinus and Kellner Weldon, Frauke and El Koussy, Marwan and Kiefer, Claus and Federspiel, Andrea and Schroth, Gerhard},
    month = sep,
    year = {2016},
    pages = {481--488},
}

@article{goyal_endovascular_2025,
    title = {Endovascular {Treatment} of {Stroke} {Due} to {Medium}-{Vessel} {Occlusion}},
    volume = {392},
    copyright = {http://www.nejmgroup.org/legal/terms-of-use.htm},
    issn = {0028-4793, 1533-4406},
    doi = {10.1056/NEJMoa2411668},
    language = {en},
    number = {14},
    urldate = {2025-06-20},
    journal = {New England Journal of Medicine},
    author = {Goyal, Mayank and Ospel, Johanna M. and Ganesh, Aravind and Dowlatshahi, Dar and Volders, David and Möhlenbruch, Markus A. and Jumaa, Mouhammad A. and Nimjee, Shahid M. and Booth, Thomas C. and Buck, Brian H. and Kennedy, James and Shankar, Jai J. and Dorn, Franziska and Zhang, Liqun and Hametner, Christian and Nardai, Sandor and Zafar, Atif and Diprose, William and Vatanpour, Shabnam and Stebner, Alexander and Bosshart, Salome and Singh, Nishita and Sebastian, Ivy and Uchida, Kazutaka and Ryckborst, Karla J. and Fahed, Robert and Hu, Sherry X. and Vollherbst, Dominik F. and Zaidi, Syed F. and Lee, Vivien H. and Lynch, Jeremy and Rempel, Jeremy L. and Teal, Rachel and Trivedi, Anurag and Bode, Felix J. and Ogungbemi, Ayokunle and Pham, Mirko and Orosz, Peter and Abdalkader, Mohamad and Taschner, Christian and Tarpley, Jason and Poli, Sven and Singh, Ravinder-Jeet and De Leacy, Reade and Lopez, George and Sahlas, Demetrios and Chen, Michael and Burns, Paul and Schaafsma, Joanna D. and Marigold, Richard and Reich, Arno and Amole, Adewumi and Field, Thalia S. and Swartz, Richard H. and Settecase, Fabio and Lenzsér, Gábor and Ortega-Gutierrez, Santiago and Asdaghi, Negar and Lobotesis, Kyriakos and Siddiqui, Adnan H. and Berrouschot, Joerg and Mokin, Maxim and Ebersole, Koji and Schneider, Hauke and Yoo, Albert J. and Mandzia, Jennifer and Klostranec, Jesse and Jadun, Changez and Patankar, Tufail and Sauvageau, Eric and Lenthall, Robert and Peeling, Lissa and Huynh, Thien and Budzik, Ronald and Lee, Seon-Kyu and Makalanda, Levansri and Levitt, Michael R. and Perry, Richard J. and Hlaing, Thant and Jahromi, Babak S. and Singh, Paul and Demchuk, Andrew M. and Hill, Michael D.},
    month = apr,
    year = {2025},
    pages = {1385--1395},
}

@article{Messe2016WhyTPA,
    title = {{Why are acute ischemic stroke patients not receiving IV tPA?}},
    year = {2016},
    journal = {Neurology},
    author = {Mess{\'{e}}, Steven R. and Khatri, Pooja and Reeves, Mathew J. and Smith, Eric E. and Saver, Jeffrey L. and Bhatt, Deepak L. and Grau-Sepulveda, Maria V. and Cox, Margueritte and Peterson, Eric D. and Fonarow, Gregg C. and Schwamm, Lee H.},
    number = {15},
    month = {10},
    pages = {1565--1574},
    volume = {87},
    doi = {10.1212/WNL.0000000000003198},
    issn = {0028-3878}
}

@article{baid2021rsna,
  title={The rsna-asnr-miccai brats 2021 benchmark on brain tumor segmentation and radiogenomic classification},
  author={Baid, Ujjwal and Ghodasara, Satyam and Mohan, Suyash and Bilello, Michel and Calabrese, Evan and Colak, Errol and Farahani, Keyvan and Kalpathy-Cramer, Jayashree and Kitamura, Felipe C and Pati, Sarthak and others},
  journal={arXiv preprint arXiv:2107.02314},
  year={2021}
}

@misc{OlivaresReboredo2025ExploringAssessment,
    title = {{Exploring Temporal Dynamics in No-Reflow Assessment}},
    year = {2025},
    author = {Olivares Reboredo, Carlos Andres and Athreya, Shreeram and Ismail, Ameera and Nael, Kambiz and Speier, William and Arnold, Corey},
    month = {12},
    doi = {10.64898/2025.12.08.25341865}
}

@article{Sano2025PredictorsCirculation.,
    title = {{Predictors of Futile Recanalization after Mechanical Thrombectomy for Embolism-Related Large Vessel Occlusion in the Anterior Circulation.}},
    year = {2025},
    journal = {Journal of neuroendovascular therapy},
    author = {Sano, Takanori and Kobayashi, Kazuto and Tanemura, Hiroshi and Ishigaki, Tomoki and Miya, Fumitaka},
    number = {1},
    volume = {19},
    doi = {10.5797/jnet.oa.2025-0068},
    issn = {2186-2494},
    pmid = {41059091}
}

@article{Kiani2025NeuroimagingMeta-Analysis.,
    title = {{Neuroimaging Predictors of Futile Recanalization in Anterior Circulation Stroke: A Systematic Review and Meta-Analysis.}},
    year = {2025},
    journal = {AJNR. American journal of neuroradiology},
    author = {Kiani, Iman and Mohammadzadeh, Saeed and Mozaffari, Sana and Mahmoudzadeh, Hanieh and Lakhani, Dhairya A and Kakadiya, Jay and Salim, Hamza A and Aziz, Yasmin N and Sriwastwa, Aakanksha and Majmundar, Shyam and Mei, Janet and Dmytriw, Adam A and Guenego, Adrien and Xu, Risheng and Lu, Hanzhang and Hillis, Argye E and Albers, Gregory W and Liebeskind, David and Shah, Gaurang V and Nael, Kambiz and Heit, Jeremy J and Faizy, Tobias D and Yedavalli, Vivek S},
    month = {11},
    doi = {10.3174/ajnr.A9095},
    issn = {1936-959X},
    pmid = {41213813}
}

@article{Sun2024No-reflowStrategies.,
    title = {{No-reflow after recanalization in ischemic stroke: From pathomechanisms to therapeutic strategies.}},
    year = {2024},
    journal = {Journal of cerebral blood flow and metabolism : official journal of the International Society of Cerebral Blood Flow and Metabolism},
    author = {Sun, Feiyue and Zhou, Jing and Chen, Xiangyu and Yang, Tong and Wang, Guozuo and Ge, Jinwen and Zhang, Zhanwei and Mei, Zhigang},
    number = {6},
    month = {6},
    pages = {857--880},
    volume = {44},
    doi = {10.1177/0271678X241237159},
    issn = {1559-7016},
    pmid = {38420850}
}

@article{Kanoke2020TheStroke,
    title = {{The impact of native leptomeningeal collateralization on rapid blood flow recruitment following ischemic stroke}},
    year = {2020},
    journal = {Journal of Cerebral Blood Flow {\&} Metabolism},
    author = {Kanoke, Atsushi and Akamatsu, Yosuke and Nishijima, Yasuo and To, Eric and Lee, Chih C and Li, Yuandong and Wang, Ruikang K and Tominaga, Teiji and Liu, Jialing},
    number = {11},
    month = {11},
    pages = {2165--2178},
    volume = {40},
    doi = {10.1177/0271678X20941265},
    issn = {0271-678X}
}

@article{Freitas-Andrade2020StructuralStroke,
    title = {{Structural and Functional Remodeling of the Brain Vasculature Following Stroke}},
    year = {2020},
    journal = {Frontiers in Physiology},
    author = {Freitas-Andrade, Moises and Raman-Nair, Joanna and Lacoste, Baptiste},
    month = {8},
    volume = {11},
    doi = {10.3389/fphys.2020.00948},
    issn = {1664-042X}
}

@article{Qiu2023ARecanalization,
    title = {{A systematic observation of vasodynamics from different segments along the cerebral vasculature in the penumbra zone of awake mice following cerebral ischemia and recanalization}},
    year = {2023},
    journal = {Journal of Cerebral Blood Flow {\&} Metabolism},
    author = {Qiu, Baoshan and Zhao, Zichen and Wang, Nan and Feng, Ziyan and Chen, Xing-jun and Chen, Weiqi and Sun, Wenzhi and Ge, Woo-ping and Wang, Yilong},
    number = {5},
    month = {5},
    pages = {665--679},
    volume = {43},
    doi = {10.1177/0271678X221146128},
    issn = {0271-678X}
}

@article{Cipolla2014PostischemicArterioles,
    title = {{Postischemic Reperfusion Causes Smooth Muscle Calcium Sensitization and Vasoconstriction of Parenchymal Arterioles}},
    year = {2014},
    journal = {Stroke},
    author = {Cipolla, Marilyn J. and Chan, Siu-Lung and Sweet, Julie and Tavares, Matthew J. and Gokina, Natalia and Brayden, Joseph E.},
    number = {8},
    month = {8},
    pages = {2425--2430},
    volume = {45},
    doi = {10.1161/STROKEAHA.114.005888},
    issn = {0039-2499}
}

\end{document}